\documentclass{emulateapj}

\catcode`\@=11
\def\gapprox{\mathrel{\mathpalette\@versim>}}
\def\lapprox{\mathrel{\mathpalette\@versim<}}
\def\@versim#1#2{\lower2.45pt\vbox{\baselineskip0pt\lineskip0.9pt
     \ialign{$\m@th#1\hfil##\hfil$\crcr#2\crcr\sim\crcr}}}
\def\HI{{\ion{H}{1}}}
\def\HII{{\ion{H}{2}}}

\catcode`\@=12
\newcommand{\hi}{H\,{\sc i}}
\newcommand{\hii}{H\,{\sc ii}}
\newcommand{\neii}{Ne\,{\sc ii}}
\newcommand{\msolar}{\ensuremath{\mbox{M}_\odot}}
\newcommand{\lsolar}{\ensuremath{\mbox{L}_\odot}}
\newcommand{\OmM}{\ifmmode {\Omega_{\rm M}}\else $\Omega_{\rm M}$\fi}
\newcommand{\OmL}{\ifmmode {\Omega_{\Lambda}}\else $\Omega_{\Lambda}$\fi}
\newcommand{\kmps}{\ifmmode {\rm\,km~s^{-1}} \else ${\rm\,km\,s^{-1}}$\fi}
\newcommand{\pscm}{\ensuremath{\,\mbox{cm}^{-2}}}
\newcommand{\pccm}{\ensuremath{\,\mbox{cm}^{-3}}}
\newcommand{\mapiii}{MAPPINGS  \textsc{iii}}
\newcommand{\fluxnu}{\ensuremath{\,\mbox{erg}\,\mbox{cm}^{-2}\mbox{s}^{-1}
        \mbox{Hz}^{-1}\mbox{Sr}^{-1}}}
\newcommand{\flux}{\ensuremath{\,\mbox{erg}\,\mbox{s}^{-1}\,\mbox{cm}^{-2}}}
\newcommand{\ddt}{\ensuremath{\frac{d}{dt}}}
\newcommand{\Msunpyr}{\ensuremath{{\rm M}_{\odot}{\rm yr}^{-1}}}
\newcommand{\Cpar}{\ensuremath{{\cal C}}}
\newcommand{\mum}{\ensuremath{\mu\mbox{m}}}
\received{}
\begin{document}

\title{Modelling the Pan-Spectral Energy Distribution of Starburst
Galaxies: IV.\\ 
The Controlling Parameters of the Starburst SED}

\author{Brent Groves\altaffilmark{1}, Michael A. Dopita\altaffilmark{2}, 
Ralph S. Sutherland\altaffilmark{2}, Lisa J. Kewley\altaffilmark{3}, 
J\"{o}rg Fischera\altaffilmark{4}, Claus Leitherer\altaffilmark{5}, Bernhard Brandl\altaffilmark{1}, \& Wil van Breugel\altaffilmark{6}}
\altaffiltext{1}{Sterrewacht Leiden, Leiden University, P.O. Box 9513, 2300 RA Leiden, The Netherlands}
\altaffiltext{2}{Research School of Astronomy \& Astrophysics,
The Australian National University, \\
Cotter Road, Weston Creek, ACT 2611, Australia}
\altaffiltext{3}{University of Hawai'i, Institute for Astronomy, 2680 Woodlawn Drive, Honolulu, HI, 96822 USA}
\altaffiltext{4}{Canadian Institute for Theoretical Astrophysics, Univ. of Toronto,
60 St. George Street, Toronto, Ontario, M5S 3H8, Canada}
\altaffiltext{5}{Space Telescope  Science Institute, 3700 San Martin Drive, Baltimore  MD21218, USA}
\altaffiltext{6}{U. Cal. at Merced, P.O. Box 2039, Merced, CA 95344, USA}
\email{brent@strw.leidenuniv.nl}


\begin{abstract}
We combine the the stellar spectral synthesis code
Starburst99, the nebular modelling 
code \mapiii\, and a 1-D dynamical evolution model of \hii\ regions
around massive clusters of young stars to generate improved models of
the spectral energy distribution (SED) of starburst galaxies. We
introduce a compactness parameter, \Cpar , which characterizes the
specific intensity of the radiation field at ionization fronts in \hii\
regions, and which controls the shape of the far-IR dust re-emission,
often referred to loosely as the dust ``temperature''.  We also
investigate the effect of metallicity on the overall SED and in
particular, on the strength of the PAH features. We provide templates
for the mean emission produced by the young compact \hii\ regions, the
older ($10 - 100$\ Myr) stars and for the wavelength-dependent
attenuation produced by a foreground screen of the dust used in our
model. We demonstrate that these components may be combined to produce
a excellent fit to the observed SEDs of star formation dominated
galaxies which
are often used as templates (Arp 220 and NGC 6240). This fit extends
from the Lyman Limit to wavelengths of about one mm. The methods
presented in both this paper and in the previous papers of this series 
allow the
extraction of the physical parameters of the starburst region (star
formation rates, star formation rate history, mean cluster mass,
metallicity, dust attenuation and pressure) from the analysis of the
pan-spectral SED. 
\end{abstract}


\section{Introduction}\label{sec:intro}

By definition, the bolometric luminosity of a starburst galaxy is
dominated by the 
young stars it contains. Thus regardless of how much or how little of
this luminosity is reprocessed through the dusty interstellar medium
either through thermal emission in the IR of dust grains, through
fluorescent processes, or through heating and re-emission in an ionized
medium, the pan-spectral SED encodes information about what the star
formation rate currently is, and what it has been in the recent
past. The first objective of pan-spectral SED modelling is therefore
to be able to reliably infer star formation rates in galaxies and to
provide likely error estimates using observational data sets which may
in practice be restricted to only certain emission lines or spectral
features. In principle, almost any part of the SED of a starburst can
be used as a star formation indicator,  provided that the appropriate
bolometric correction to the absolute luminosity can be made, and
observational issues are accounted for. In practice, each wavelength
regime has a different level of sensitivity to the ongoing star
formation which is dependent upon these bolometric corrections, 
with hydrogen
emission lines and the IR part of the SED being the most robust
indicators of the current star formation rate (SFR). These
bolometric corrections critically depend on the foreground dust
absorption (more properly called dust attenuation), and the geometry
of the embedded dusty molecular clouds, with respect to both the
ionizing stars and the older stellar population. The accurate
determination of such bolometric corrections is a major motivation of
our theoretical work of pan-spectral SED modelling. 

In order to correctly model the  SEDs of starburst galaxies, we
first need to understand how the form of the SED is controlled by the
interstellar physics and the geometry of the stars with respect to the
gas. Once these are understood, we can then use our theoretical models
to attain the objectives of our second motivation for such SED
modelling; to gain insights into the physical parameters of starburst
galaxies. In particular, we can hope to quantify the stellar
populations, the atomic and molecular gas content, the star and
gas-phase metallicities, physical parameters  of their interstellar
media such as the pressure or mean
density, and the nature of the
interstellar dust, both its composition and spatial distribution. The
physical parameters so derived on homogeneous samples of objects can
then help develop our insight into the physical processes which
control them.  

The dust grain temperature distribution, and therefore the shape and
peak of the far-IR feature, depends critically on the geometrical
relationship between the dust grains and the stellar heating sources
assumed in the model. Models with warmer far-IR colors will have a
more compact disposition of gas with respect to the stars. The
difficulty here is that, in any simple starburst model, these
geometrical relationships are not determined \emph{a priori}.  

In the semi-empirical modelling of Dale and his collaborators
\citep{Dale01, Dale02}, the SEDs of both disk and starburst galaxies
were suggested to form a one-parameter family in terms of dust
temperature. This suggested that starburst galaxies have hotter dust
temperatures. \cite{Lagache03} (again empirically) have suggested that
the absolute luminosity controls the form of the SED. Both of these
assertions may be true to some extent, since IR luminous galaxies
generally have greater rates of star formation than normal galaxies. 

The French group \citep{Galliano03} take the simplest approach of
approximating the starburst by a spherical \HII\ region and clumpy
dust shell around the central star forming region. A more advanced
approach is used by \citet{Siebenmorgen07} to model starbursts and
ULIRGs. While assuming a spherical geometry for the radiative
transfer, they also include the effect of the hot dust around young
OB-stars as well as the diffuse ISM dust surrounding the starburst and
older stellar population. The hot dust component is important as it
can dominate the mid-IR  emission \citep{Krugel94}. 

Associated with the latter models is the approach by
\citet{Efstathiou00}. Likewise concentrating on starburst galaxies,
they modelled the starburst as a group of stars surrounded by thick
molecular clouds in a manner similar to our approach, and in addition,
also included a similar, simple description for the evolution of the
distance of the molecular clouds to the illuminating stars. With these
models they could explain the observed IRAS distributions and
reproduce several ISO observations.

One of the more
sophisticated approaches is taken in the GRASIL code by the
Padova--Trieste group \citep{Silva98,Granato00}. Their
starburst model uses a spherical geometry with King profiles, and they
allow for the formation of clusters of stars in molecular complexes,
and their subsequent escape from these regions. This group has since
incorporated gas physics by use of the CLOUDY code
\citep{Ferland98,Abel05} to 
provide emission line diagnostics as well as dust continuum
diagnostics \citep{Panuzzo03}. 

A similarly advanced approach was used by \citet{Piovan06a,Piovan06b},
who also assume a spherical geometry with King profiles and young
stars in molecular complexes. 

In the conceptually sophisticated models of 
\citet{Takagi03a, Takagi03b}, a mass-radius relationship for the star
formation region of $r_i/{\rm kpc} = \Theta (M/{10^9 {\rm
    \msolar}})^{1/2}$ is adopted along with a stellar density
distribution given by a generalized King profile. The parameter $
\Theta $ is a compactness parameter which expresses the degree of
matter concentration, and is related to the optical depth of the dust
through which the starburst region is seen. For a sample of
ultra-luminous starbursts, they find that, while most conform to a
constant surface brightness of order $10^{12}\lsolar$kpc$^{-2}$, there
are a few objects with surface brightnesses roughly ten times larger
than this, which they ascribe to post-merger systems. In this paper,
we will adopt a derivative version of this concept of a compactness
parameter as the factor which provides the main control on the shape
of the far-IR bump. 

All the fully theoretical (as opposed to semi-empirical) methods used
by other groups involve the calculation of essentially a single
spherical radiation transfer problem. Unfortunately, real starburst galaxies
have many separate clusters of many different ages distributed in a
complex spatial distribution. 
However, gas column densities and
pressures can be extremely high, and as a consequence the size scale
of individual \HII\ region complexes can be extremely small in
comparison to the overall scale of the starburst. It is only when the
star forming complexes join up to produce large-scale collective
phenomena as in, for example, the outflow in M82, that we need to go
to a full 3-D radiative transfer model covering the whole galaxy.  

In the earlier papers in this series, we take advantage of the
localized radiative transfer approximation to construct our
pan-spectral SEDs. Instead of treating the starburst as a single \HII\
region complex covering the whole starburst region, we split the
starburst up into many individual \HII\ regions, each ionized by the UV
photons of the clusters within them, and each evolving in radius and
internal pressure according to the mechanical energy input of the
exciting stars through their stellar winds and supernova
explosions. The global SED is then the sum of the SEDs produced by
each of these \HII\ regions and their surrounding photo-dissociation
regions (PDRs) integrated over all cluster ages. Since the radiative
transfer problem in each \HII\ region is fully treated, in this
approach we stand a better chance of capturing the full range of
physical conditions encountered in a starburst region. This collective
approach is what can be found in many of the sophisticated modelling
codes such as that of \citet{Piovan06a} and GRASIL \citep{Silva98}. In
GRASIL for example, multiple core molecular cloud systems can be
included with
differing parameters for each, such as mass and optical depth. 

In paper I \citep{SED1}, we investigated the role that pressure alone
plays in changing the compactness of the \HII\ regions within the
starburst and hence in the controlling the shape of the far-IR dust
emission bump. However, a defect in these models is that they were
only run with a single value of mean cluster mass $\left<M_{\rm cl}
\right>$. A change in cluster mass directly affects the specific
intensity of the radiation field in the \HII\ region, and this will in
turn change the shape of the far-IR bump. 

In papers II \& III \citep{SED2, SED3}, we introduced the ${\cal
R}=\left<M_{\rm cl} \right>/P_{0}$ 
parameter which controls the absolute value of the ionization
parameter in the \HII\ region and its time evolution. The ionization
parameter is defined as the ratio of the ionizing photon density to
the particle density in the \HII\ region;  ${\cal U} = L_{\rm UV}/4\pi
R_{\rm HII}^2nc$, where $L_{\rm UV}$ is the flux of ionizing UV
photons produced by the central cluster, $R_{\rm HII}$ is the mean
radius of the \HII\ region with particle density $n$, and $c$ is the
speed of light. All models having a given value of ${\cal R}$ and
metallicity $Z$ will show the same run of ionization parameter as a
function of cluster age and will therefore produce identical line
ratios at any given age. 

The shape of the dust feature or ``bump'' in the IR in starbursts is
controlled by the distribution of dust temperatures in the starburst
galaxy. Within a given \HII\ region, this distribution of temperatures
is controlled by the specific photon density, meaning that the mean dust
temperature of any individual grain is $\left<T_{\rm gr} \right> = f(L_{\rm
  UV}/R_{\rm HII}^2)$. Thus denser and more compact \HII\
regions will produce hotter grain temperature distributions. In this
paper, we introduce, by analogy with the ${\cal R}$ parameter, a
``Compactness Parameter'', ${\cal C}$. All models having a given value
of ${\cal C}$ and metallicity $Z$ will show the same run of grain
temperature distribution as a function of cluster age and will
therefore produce identical far-IR dust re-emission bumps at any given
age. 

In the following sections of this paper, we will discuss the details
of our modeling procedure, insofar as this is different from that used
in earlier papers in this series, introduce the Compactness Parameter
and show that this does indeed serve to characterize the far-IR bump
for \HII\ regions in the age range $0.5 - 10$\ Myr. We also investigate
the effects of varying metallicity on the form of the far-IR bump,
provide templates for the mean SED of compact \HII\ regions, and of
the older ($10 - 100$\ Myr) stars, and for the attenuation produced by a
dusty fractal foreground screen. Finally, we show how these components
can be combined to produce excellent fits to frequently-used starburst
templates such as Arp~220 or NGC~6240. 

\section{Models}

The Starburst models calculated here follow the general form described
in the previous papers of this series \citep{SED1,SED2,SED3};
hereinafter SED1, SED2 and SED3, respectively. However, apart from
being updated to use the latest versions of the modelling codes
Starburst99 \citep{Leitherer99} and \mapiii\ \citep{Groves04}, the
models incorporate a number of changes or improvements that we here
describe in greater detail. To make this paper self-contained, we
briefly recapitulate on the techniques used in the earlier papers of
this series. 

\subsection{Stars and Stellar Clusters}\label{sec:stars}

We have used the latest version (2006) of
Starburst99\footnote{http://www.stsci.edu/science/starburst99/}
to compute the spectral energy distribution of clusters of stars of
any given age. A 
detailed description of latest stellar atmospheres and stellar
evolution physics used within the code are given in \citet{Smith02}
and \citet{Vazquez05}. 

In our Starburst99 models, we take an instantaneous burst of $M_{\rm
  cl}=10^{6} M_{\odot}$, having a \citet{Kroupa02} broken power-law
IMF between 0.1 and 120 $M_{\odot}$. Within the code we use the
standard combination of the Geneva and Padova tracks for the stellar
evolution \citep{Vazquez05}, and these determine the total mechanical
luminosity, $L_{\rm mech}$. We use the theoretical ``high-mass-loss''
tracks for the treatment of the stellar wind.  The supernova cutoff
mass is 8 $M_{\odot}$. However, since the \HII\ region evolution is
run up to only 10\ Myr, the exact choice of this cutoff mass is
unimportant to the modelling.   

We output the stellar spectra at 0.01\ Myr, 0.5\ Myrs and then
every 0.5\ Myrs after that up to an age of 10\ Myrs, by which effectively all
of the ionizing photons have been emitted (see SED2). Note that
the resolution of the starburst model is actually higher at 0.1\ Myrs,
which provides the fine-gridding needed to accurately track the
mechanical energy input used in the computation of the evolution of
the \HII\ regions. For completeness, this model was also extended to 1
Gyr, with spectra computed at longer intervals for older stellar
templates. 

To determine the parameters for clusters of any given mass, we assume
a simple scaling with cluster mass. Such a scaling should hold in
starbursting regions where many clusters are forming, as here 
stochastic effects are generally small, and on average the IMF is well
sampled throughout the mass range. However, as a number of authors
have pointed out, several of whom are referenced in SED2, and
most recently by \citet{Weidner06}, such an assumption will not hold
for low cluster masses.  

\subsection{\hii\ Region Evolution}\label{sec:hiievol}

The \HII\ regions are treated as 1-D mass-loss bubbles driven by the
mechanical energy input of their stars and supernovae
\citep{Castor75}. Their equation of motion is given by (SED1): 

\begin{equation}\label{eqn:Rdiff}
\ddt\left[R\ddt\left(R^{3}\dot{R}\right)\right] +
\frac{9}{2}R^{2}\dot{R}^{3} = \frac{3L_{\rm mech}(t)}{2\pi\rho_{0}},
\end{equation}
where the time dependent mechanical luminosity, $L_{\rm mech}(t)$, is
determined from the Starburst99 output. 

The pressure in the \hii\ region with radius $R$, expanding in an ISM
with density $\rho_{0}$ is then determined as, 
\begin{eqnarray}
P&=& n_{\rm HII}kT_{e}, \nonumber \\
&=& \frac{7}{(3850\pi)^{2/5}}\left(\frac{250}{308\pi}\right)^{4/15}
\left(\frac{L_{\rm mech}(t)}{\rho_{0}}\right)^{2/3}
\frac{\rho_{0}}{R^{4/3}(t)}, 
\end{eqnarray}\label{eqn:hiiP}
where $n_{HII}$ is the density in the ionized gas, with electron
temperature $T_{e}$. This equation is derived from the
\citep{Oey97,Oey98} version of the \citet{Castor75} mass-loss bubble
formulae with  the assumption that the \hii\ region has the same
pressure as the shocked stellar wind, and is confined to a thin shell
around the periphery of the wind-blown bubble. The ionizing flux at
the inner boundary of the \hii\ region is then  $L_{UV}/4\pi R^2$ and
the density in the ionized region  $n_{HII}$ is given by equation
\ref{eqn:hiiP}, from which the ionization parameter of the \HII\
region, $\cal{U}$,  can be derived. 

\subsection{Nebular Abundances and Depletion Factors}\label{sec:abund}

The abundance set and depletion factors used in these models are
unchanged from those presented in SED3 and are given in table
\ref{tab:abund}. The nebular abundance set follows \citet{Asplund05}.  
As noted previously, the gas phase ``Solar'' abundance in the models
is somewhat offset from 
the ``Solar'' abundance set used in Starburst99. While this has
been shown to have no significant effect on the models, it does result
in a small inconsistency in the models.

The exploration of metallicity effects within the models presented
here is limited to those metallicities computed in Starburst99;
$0.05Z_{\odot}$, $0.2Z_{\odot}$, $0.4Z_{\odot}$, $1.0 Z_{\odot}$, and
$2.0Z_{\odot}$. As in SED3, we simply scale the abundances with
metallicity, with the following exceptions. For helium we use the
empirical relationship to include the primordial component as well as
that from nucleosynthesis: 
\begin{equation}
{\rm He}/{\rm H} = 0.0737 + 0.024(Z/Z_{\odot}).
\end{equation}
The elements carbon and nitrogen  are both observed to have
a primary nucleosynthetic component and a secondary nucleosynthetic component at
higher metallicities. Thus for these elements we adopt the following
empirical relationships;
\begin{eqnarray}
{\rm C}/{\rm H} & = & 6.0\times10^{-5}(Z/Z_{\odot}) +
2.0\times10^{-4}(Z/Z_{\odot})^2 \\
{\rm N}/{\rm H} & = & 1.1\times10^{-5}(Z/Z_{\odot}) +
4.9\times10^{-5}(Z/Z_{\odot})^2 
\end{eqnarray}

The depletion factors used in the models
are based upon the observed depletion fractions in the local interstellar cloud
\citep{Kimura03}. These may not represent in reality those found in the
starburst environments, but currently we have no adequate way
of estimating these. We must also assume that the depletion pattern
does not vary with 
metallicity, as there are currently no good models or observations for the 
relationship between metallicity and depletion. However, in the
starbursting environments modelled here, such an assumption may be
adequate \citep{Draine07}. With this assumption the
dust-to-gas ratio is purely a function of metallicity.

\subsection{Photoionization models}\label{sec:models}

For the component \hii\ regions, we compute both the emission and
internal absorption of both gas (line  
plus continuum) and dust (continuum including polycyclic aromatic
hydrocarbons) using  
the \mapiii\ code, with the Starburst99 stellar cluster
spectra as input. 

Apart from the effects of burst age and metallicity, here we explore
three other parameters within the models; the ISM pressure, $P_{0}$,
the cloud covering fraction, $f_{\rm PDR}$, and the Compactness
Parameter, ${\cal C}$, described below, which is a
function both of $P_{0}$ and of the mean cluster mass, $\left< M_{\rm
  cl} \right >$.  

For the models investigated here we examine five different thermal gas
pressures;
$P_{0}/k=10^4,~10^5,~10^6,~10^7,$ and $10^8$ K \pccm. These five
pressures cover the full range expected to be encountered in
starbursting galaxies, from regions of enhanced star formation in disc
galaxies, to the high pressure ULIRGs. 

For each parameter set we compute models at 21 starburst ages,
covering the timescale 0.01\ Myrs to 10\ Myrs in steps of 0.5\ Myrs.  

Finally, for each age we run two models. The first model is of the
\hii\ region alone;  the region within which 99\% of the hydrogen line
emission arises, and within which almost all of the ionizing photons
are absorbed. It is within this region that the hottest dust emission
arises. The second model corresponds to the Photodissociation Region
(PDR) surrounding the \HII\ region. This region is the transition
layer between the \hii\ region and surrounding dense molecular cloud, from
which the stellar cluster is thought to have formed.
In the PDR,  a large
fraction of the stellar light is absorbed and most of the PAH and dust
far-IR emission produced. We follow the radiative transfer beyond the
ionization front in the \HII\ region until a total hydrogen column
depth of $N({\rm HI})=10^{22}$\pscm\ is achieved.  The column depth of
$10^{22}$\pscm\ is based upon both observations and upon theoretical
considerations. The observational data comes from measurements of
individual molecular clouds both within our galaxy \citep{Larson81,
  Solomon87, Heyer01} and in neighbouring galaxies such as M33
\citep{Rosolowsky03} which all give hydrogen column densities of
$\sim10^{22}$\pscm\, independent of cloud radius. This value is not
unexpected, since it indicates that all giant molecular clouds are
marginally stable against gravitational collapse, provided that their
virial temperatures are a few tens of degrees K. 

These two models, \hii\ and PDR, are combined through our final
parameter, $f_{\rm PDR}$, the starburst cloud covering fraction. This
parameter is a simplified version of the clearing timescale introduced
in Paper I, and discussed in other dust models
\citep[eg][]{Silva98,Charlot00}. We introduce this parameter because a
starbursting system will be a conglomerate of bursts of different ages
and sizes, unlike a single molecular cloud around an individual
cluster. Thus, while the molecular cloud clearing timescale offers a
more physical picture for an individual cluster, the starburst cloud
covering fraction better represents what is likely to be encountered
in a starbursting system. 

\begin{figure}
\includegraphics[width=0.9\hsize]{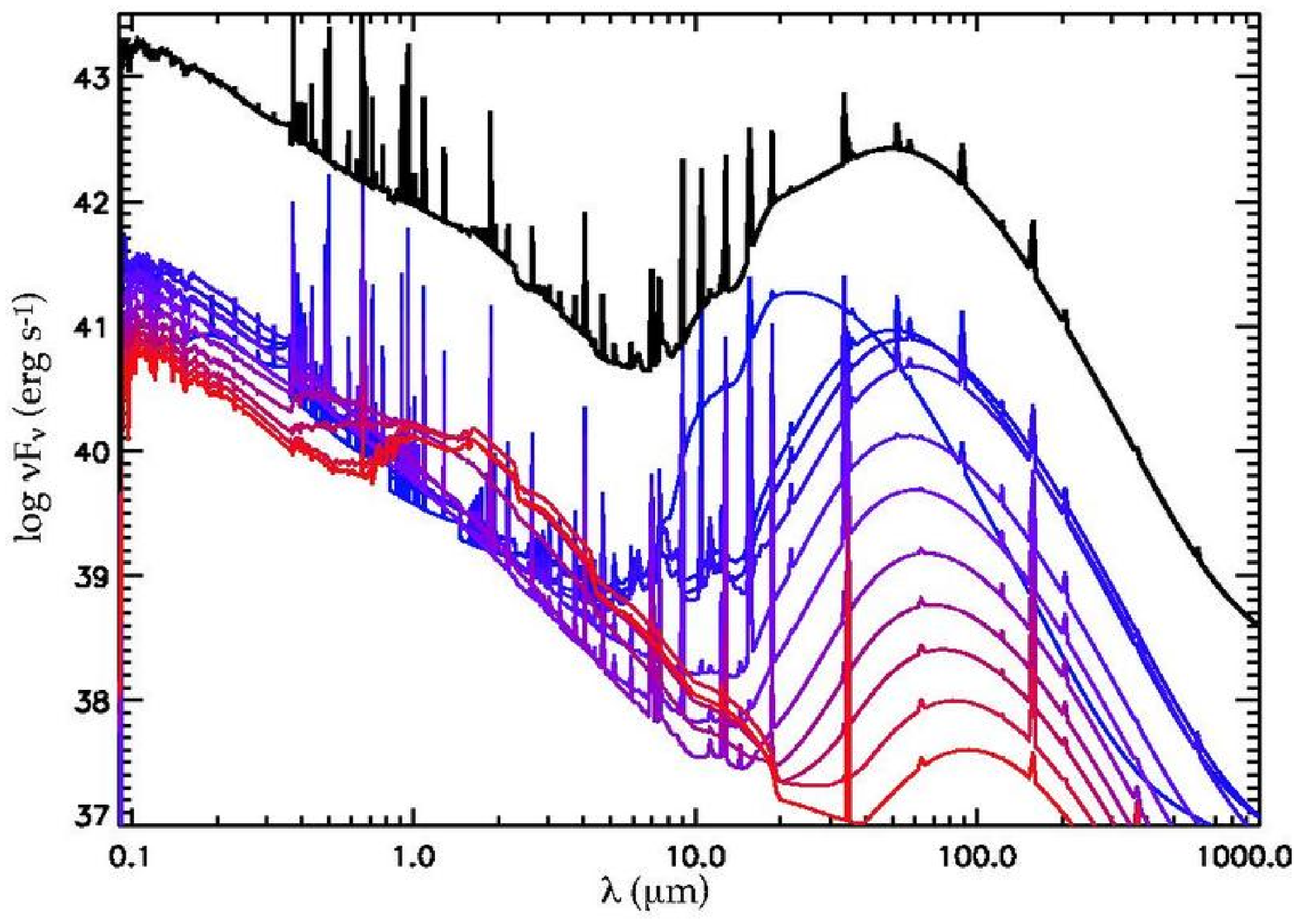}
\includegraphics[width=0.9\hsize]{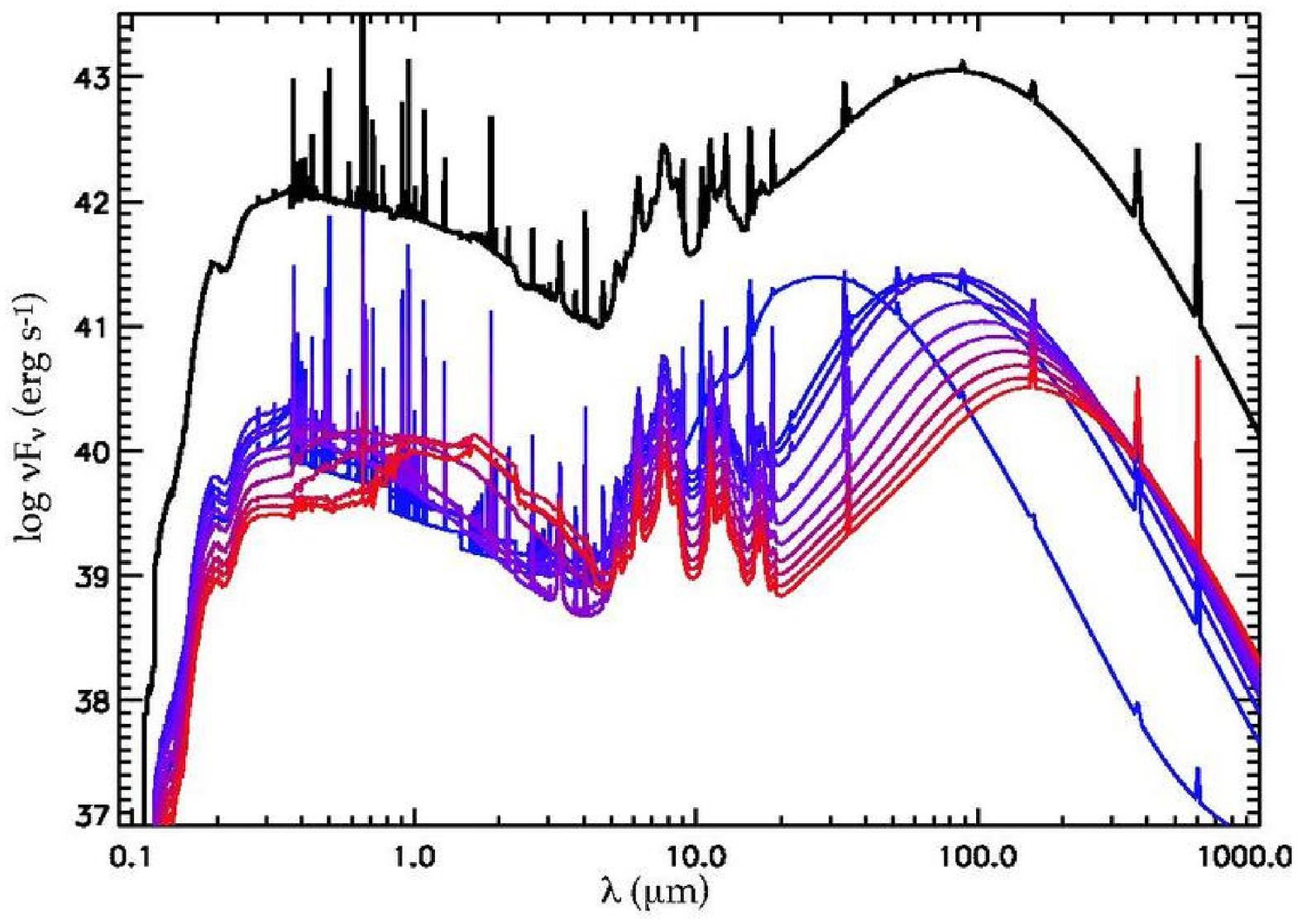}
\caption{The evolution of the spectral energy distribution with age (0
  - 10\ Myr, every 1\ Myr) for a $Z=1Z_{\odot}$, $\log {\cal C}=5.0$ 
($\log \left<M_{\rm cl}\right>=5$), $\log  P/k=5.0$
  K\pccm\ \hii\ region-only model (top) and \hii\ region plus PDR
  model (bottom).  In both diagrams, the final, integrated starburst SED is
  shown as the thick line at high luminosity.}\label{fig:Hiiages}  
\end{figure}

The multiple star clusters forming in a starburst will be of all
possible masses, and, sampled at any instant in time, of all possible
ages. To account for this we compute the luminosity-weighted average
of 
all twenty one ages computed between $0-10$\ Myr.  In figure
\ref{fig:Hiiages} we show the SEDs of the 21 calculated ages 
of an individual \hii\ region and
a \HII\ region with its PDR, along with the summed
final average SED for each. These figures clearly show the evolution
of both the stellar and nebular spectrum with age, and reveal the
decreasing cluster UV flux and cooling dust temperatures as the
clusters age and the \hii\ bubble expands. 
 
In all, we have computed  a total 300 starburst \HII\ region models
covering 5 metallicities, 6 values of the ${\cal C}$ parameter
(described below), 5 values of the ISM pressure, and a separate \hii\
region and \HII\ region plus PDR model for each set of parameters
(corresponding to $f_{\rm PDR}=0$ and $1$, respectively). All models
are scaled in flux to correspond to a star formation rate of 1
\Msunpyr\ continued over the 10\ Myr lifetime of the \HII\ regions.  

\subsection{Dust Physics}\label{sec:dust} 

The treatment of dust within the photoionization code \mapiii\ was
discussed in SED1.  Here, we concentrate only those areas where
changes have been made to the dust parameters within the code.  

In brief, our dust model consists of 3 components; Graphites, Silicates
and Polycyclic Aromatic Hydrocarbons. The optical data for each of
these comes from the series of work by Draine et
al.\footnote{http://www.astro.princeton.edu/$\sim$draine/dust/dust.diel.html
}\,\citep{Laor93,Li01}. The IR spectrum arising from dust, excluding the
effects of PAH emission, is 
calculated self-consistently, including the effects of stochastic
heating in small grains  \citep[all dust calculations are discussed in
  detail in][]{Groves04}.  The total dust-to-gas ratio within the code
is set by the fraction of metals depleted from the gas onto dust,
given in table \ref{tab:abund}.  

The heavy elements removed from the gas phase are distributed between
the two main types of dust, carbonaceous and siliceous, with the
carbonaceous dust being further divided into graphite and PAHs. The
graphite and silicate dust is distributed across a grain size
distribution arising from grain destruction processes (SED1);  
\begin{equation}
dN(a)/da = k  a^{-3.3}
\frac{e^{-(a/a_\mathrm{min})^{-3}}}{1+e^{(a/a_\mathrm{max})^3}},
\label{eqn:GSprof}
\end{equation}
with $k$ defined by the dust-to-gas ratio.

The minimum grain size of graphite and silicates are $a_{\rm
  min}=$20\AA\ and 40\AA\  respectively while the maximum grain size
is the same for both species at $a_{\rm max}=1600$\AA. 

\subsection{PAHs}
PAHs are treated somewhat separately to the other types of dust. They
are given a characteristic size and an opacity similar to coronene,
the best studied catacondensed PAH. We have constructed an
empirically-based IR emission spectrum described in more detail below.
The PAH-to-gas ratio is defined through the PAH-to-Carbon dust ratio,
which is set at 30\% in these models. While this may not be accurate
for all environments and metallicities \citep{Draine07}, it provides a
reasonable match to current observations of nearby starforming
galaxies. Differences between the template IR emission spectrum and
those actually observed will provide limits on parameters such as
de-hydrogenation, the relative abundance of cata-condensed and
peri-condensed species, and the degree of nitrogen substitution within
the carbon skeleton, which affects the 6.2 \mum\ C--C stretch feature
\citep{Peeters02}. 

There is now a great deal of observational evidence that PAHs are
destroyed within the ionized parts of the \hii\ region complexes, with
\emph{Spitzer} observations of galactic \hii\ regions showing clear
boundaries between the outer PDR PAH emitting zone and the inner
photoionized zone \citep[eg][]{Churchwell06,Povich07}. The exact
destruction mechanism is uncertain, but is likely to be
photo-destruction through stochastic heating and/or photoionization
and dissociation. To simulate this process 
within the \mapiii\ code, we previously introduced the Habing
photodissociation parameter, ${\cal H}=F_{\rm FUV}/n_{\rm H}c$, a FUV
analogy of the standard dimensionless ionization parameter ${\cal U}$  
(see eqn 17 and associated section in SED1). In a series of
test models, we found that for typical, solar metallicity starbursts,
${\cal H}\sim 10^{-3}$ at the ionization front. Henceforth we assume
this value to be the destruction point for PAHs within our models. For
typical \hii\ densities of $10-100$\pccm, this implies a radiation
field $\sim2-20$ times the \citet{Habing68} local interstellar
radiation field, consistent with the range up to which PAHs are
observed to survive \citep{Compiegne07}. 
At values of ${\cal H}< 10^{-3}$, PAHs exist in either neutral
or singly-charged states, are heated by the diffuse FUV/Optical
field and emit in the classic PAH mid-IR bands. 

This emission spectrum is determined by the natural modes of
vibration, bending and other deformations of  the planar carbon
skeleton. This spectrum is dependent upon the size and the electric
charge state of the molecule, and is modified by the effect of
non-hydrogenic end groups, including simple dehydrogenation and
skeletal atomic substitution \citep{Peeters02}.   

Thanks to the advent of space borne IR observatories, the PAH emission
spectrum has 
now been observed in many galaxies and situations, ranging from
beautiful maps of Galactic \hii\ regions \citep{Churchwell06}  to
detailed mapping of the features in both Starburst galaxies
\citep{Beirao07} and QSOs \citep{Schweitzer06}  

Given the wide ranges of possible molecular forms, it is surprising
that the form of the PAH emission spectrum in the Mid-IR is so similar
between different regions and galaxies \citep{Brandl06}. Only small
variations in the relative strengths of the PAH features have been
observed within our own Galaxy and in nearby galaxies \citep{Smith07}. 

The accuracy of using a PAH template to represent the series of bands
in the mid-IR can be estimated from the study of the variation of
these bands in nearby galaxies by \citet{Smith07}. They find on
average variations of a factor of two around the mean ratios of the
different PAH bands (their table 7), with the most significant
differences occurring in galaxies hosting weak AGN (such as LINERs). 
This suggests that our PAH template is accurate to about this factor,
with significant variations indicating differences such as PAH
ionization state, or correspondingly, the presence of a weak AGN in
the starburst galaxy. The differences between our models and the
observed PAH bands could therefore be used as diagnostics of ISM
physics, or host nuclear properties.

As discussed in SED1, we parameterize the template using a sum of
Lorentzian profiles. The Lorentzian fits to the spectrum take the
form: 
\begin{equation}
F_{\nu}\left( x\right) =\frac{f_{0}}{\left[ 1+\left(
x-x_{0}\right) ^{2}/\sigma ^{2}\right] }, \label{eqn:lorentz}
\end{equation}
where $x=1/\lambda $ cm$^{-1}$, the central wavenumber of the feature
is $x_{0}$, the  $\mathrm{FWHM}=2\sigma$, and the peak value is
$f_{0}$ (\fluxnu).

\begin{figure}
\plotone{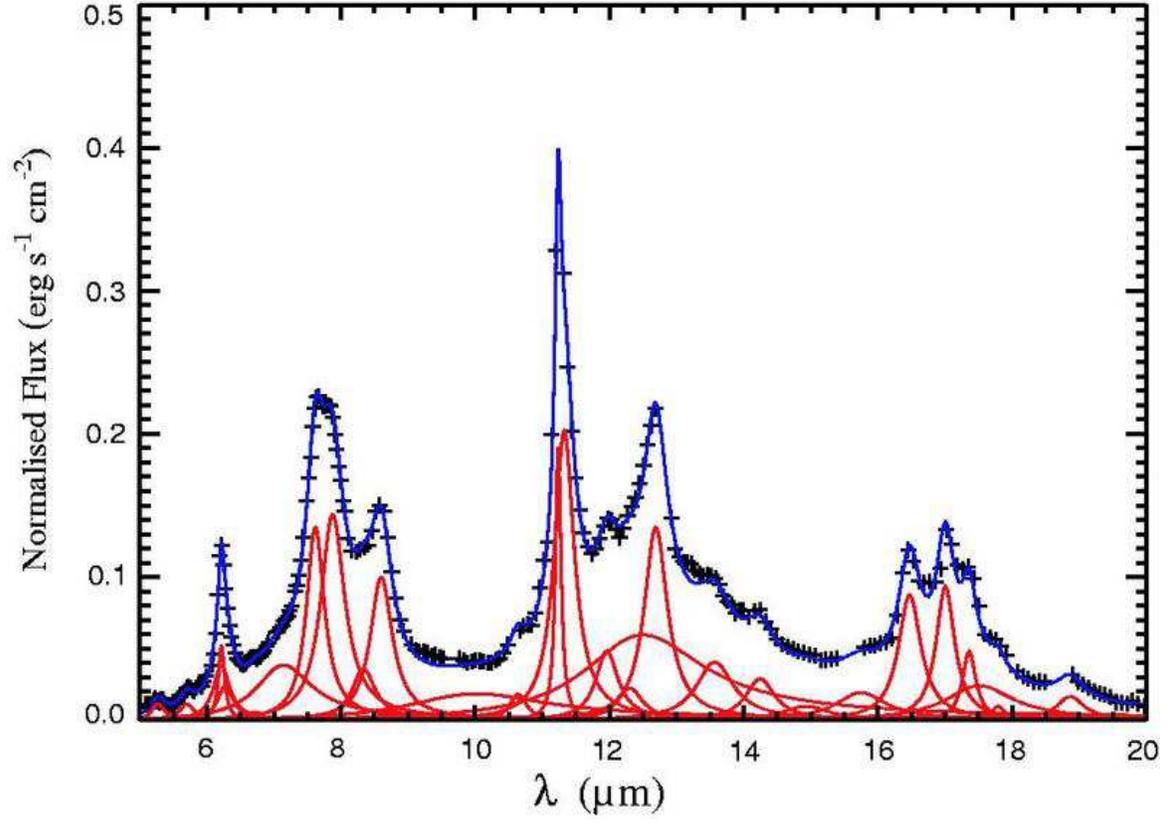}
\caption{PAH emission template. Crosses indicate the observed points,
with the solid line indicating our template fit, and the underlying
curves showing the individual Lorentzian components.}\label{fig:PAHtemplate}
\end{figure}

To derive the PAH emission spectrum template currently used in these
\mapiii\ models, we have fit Spitzer IRS observations of NGC4676 and
NGC7252. These two interacting galaxy pairs show strong, clear PAH
emission, making them good choices for a template spectrum. In both
objects we subtract 
the underlying dust continuum assuming a combined power law and
exponential form to fit the PAH-free, long wavelength end of the
spectra. The combined, continuum-subtracted observed spectrum is shown
with our best fitting template in figure \ref{fig:PAHtemplate}. The
corresponding parameters for each of the Lorentzian profiles are given
in table \ref{tab:PAHparam}.

As discussed in SED1, PAH emission within the \mapiii\ code is treated
as an energy conserving process. In equilibrium all the energy gained
by a PAH through the absorption of photons can be 
either lost through photoelectric processes or IR emission. Once the
fraction of the energy lost through 
photoelectric processes is determined using the photoelectric
cross-sections, the remaining energy fraction is re-emitted in the IR
according to our empirical template.  This is not an exact treatment,
as the PAH molecules will likely undergo stochastic heating processes
and will lose some of their energy through IR continuum emission
instead of \emph{via} these fluorescent bands. However  the code does
allow for the stochastic treatment of both very small graphite and
silicate grains. 

\section{The ${\cal C}$ Parameter}\label{sec:compact}

We now deal with the parameter which controls predominantly the form
of the far-IR 
continuum. Fundamentally, this continuum is a function of the
probability distribution of the grain temperatures throughout the
starburst. At any point within an \HII\ region or its surrounding PDR,
this is determined by the intensity and the spectral energy
distribution of the stellar radiation field. Thus, at radius $R$ in a
spherical nebula for any individual grain species, $s$: 
\begin{equation}
\left < T_{\rm gr}(s) \right > = f \left( L_*/4 \pi R^2, \bar{\nu}\right),
\end{equation}

where $\bar{\nu}$ is the mean photon energy of the radiation field,
which depends both upon the age of the cluster and the solution of the
radiative transfer problem out to radius $R$. Thus, if we wish to find
a variable in which \HII\ region models evolve along a unique grain
temperature distribution in time, $ \left < T_{\rm gr}(s,t) \right >
$, these models must also preserve the run of $L_*(t)/4 \pi
R(t)^2$. Then, because all models of this kind give a similar run of
grain temperature distributions, they will also produce very similar
global far-IR dust emission distributions. With a particular choice of
cluster luminosity, physically denser \HII\ region models have smaller
radii, and hence have hotter dust temperature distributions. 

We therefore define a \emph{Compactness Parameter}, $\cal C$, in the form:
\begin{equation}
{\cal C} \propto \frac{\left< L_{*}(t) \right>}{\left<
R(t)^2 \right>}. 
\end{equation}
This variable is akin to the compactness factor found in the models of
\cite{Takagi03a} and \cite{Takagi03b}, in that it directly determines
the dust grain temperature distribution. The ratio on the right is also 
comparable to the ratio
$m_{\rm mc}/r^2_{\rm mc}$ discussed in \citet{Silva98}, which controls
the SED of their molecular clouds and therefore the resulting hot dust
SED of their models.  
The stellar luminosity
$L_*(t)$ scales with the cluster mass $M_{\rm cl}$, and the radius and
pressure at any instant scale according to the simple mass-loss bubble
approximation of \citep{Castor75}; 
\begin{eqnarray} 
R&=&\left(\frac{250}{308\pi}\right)^{1/5}L_{\rm mech}^{1/5} 
  \rho_{0}^{-1/5}t^{3/5}\label{eqn:R_t}\\ 
P&=&\frac{7}{\left(3850\pi\right)^{2/5}}L_{\rm mech}^{2/5} 
  \rho_{0}^{3/5}t^{-4/5},\label{eqn:P_t}
\end{eqnarray}
where $L_{\rm mech}$ is the instantaneous mechanical luminosity of the
central stars of the burst (which can be assumed to scale as the mass
of the cluster) and $\rho_{0}$ the density of the ambient medium. Note
that the ambient number density is $n_{0}=\rho_{0}/(\mu m_{\rm H})$
and ambient pressure $P_{0}=n_{0}kT_{0}$.  

From these equations it follows that,
\begin{equation}
{\cal C} \propto \frac{\left< L_*(t) \right>}{\left< R(t)^2 \right>} 
\propto \frac{L_{*}}{L_{\rm mech}^{2/5}P_0^{-2/5}} 
\propto M_{\rm cl}^{3/5}P_{0}^{2/5}.
\end{equation}

This product  remains to be normalized. By analogy with the $\cal R$
parameter introduced in SED3, we choose to adopt the following
normalized definition of the Compactness Parameter; 
\begin{equation}\label{eqn:Cparam} 
\log{\cal C}=
\frac{3}{5}\log\left[ \frac{M_{\rm cl}}{M_{\odot}}\right] +
\frac{2}{5}\log\left[ \frac{P_{0}/k}{{\rm cm}^{-3}{\rm K}} \right]
\end{equation}
where $M_{\rm cl}$ should now be understood as the mean (luminosity
weighted) cluster mass in the starburst. The likely allowable physical
range on $\log{\cal C}$ in starburst environments  is, roughly, 3 -
7.5. These extremes correspond to $\log\left [ {{(P/k)} / {\pccm{\rm
K}}} \right ] \sim 4$ and $\log\left[ {{M_{\rm cl}} 
/ {M_{\odot}}}\right ]  \sim 3.5$ and  $\log\left [ {{(P/k)} /
{\pccm{\rm K}}} \right ]  \sim 8$ and $\log\left[
{{M_{\rm cl}} / {M_{\odot}}}\right ]  \sim 7$, respectively. 

\begin{figure}
\plotone{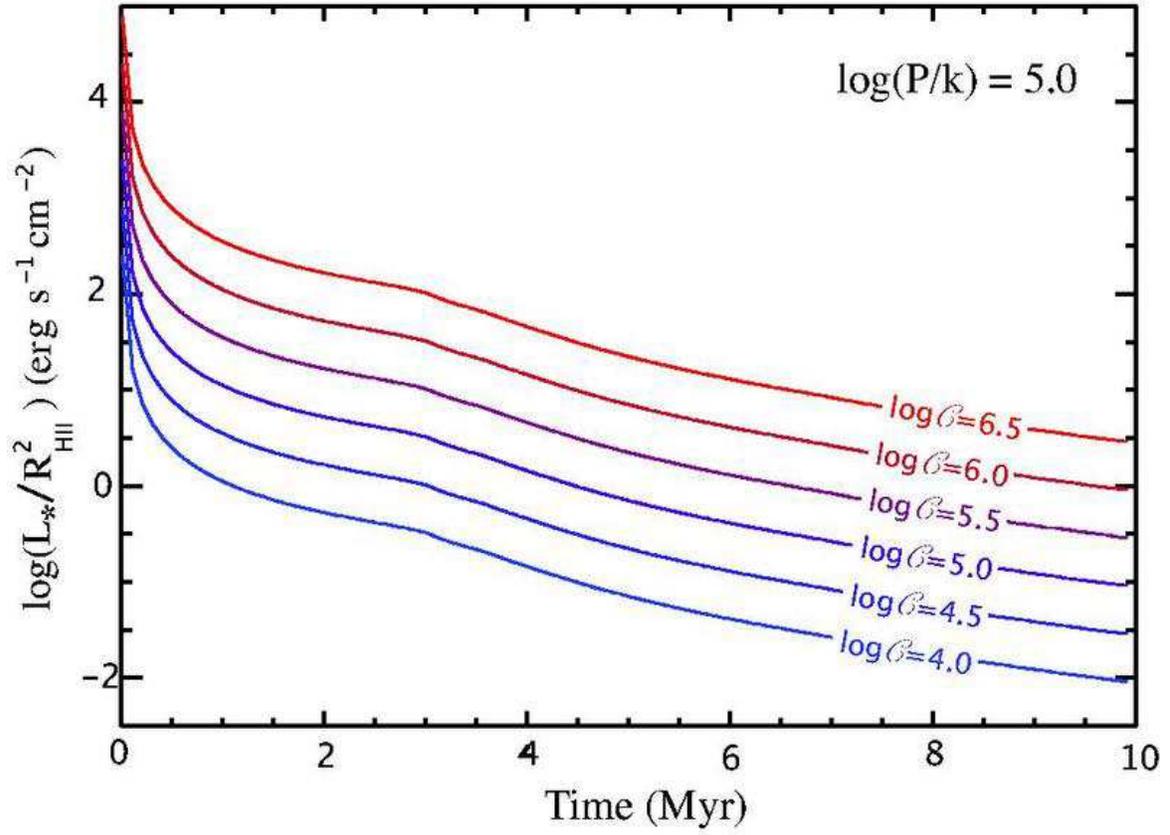}
\caption{The time variation of the incident heating flux
  ($L_{*}/R^{2}_{\rm HII}$) entering the \hii\ region surrounding a
  solar metallicity starburst with $\log(P_{0}/k)=5$. Each curve
  represents a different $\Cpar$ ($M_{\rm cl}$), and reveals how the
  \Cpar\ parameterizes different $L_{*}/R^{2}_{\rm HII}$ at any given 
  time.}\label{fig:fluxvstime}
\end{figure}

In figure \ref{fig:fluxvstime}, we show the variation of
$L_{*}/R^{2}_{\rm HII}$ with time for six different \Cpar\
parameters. These display a strong decrease of the incident flux, and
therefore of dust temperature with time. The specific flux changes
several orders of magnitude over 10\ Myrs. This decrease is due to both
the increase of the \hii\ bubble radius with time (see fig.~1 in
SED1), and the decreasing cluster luminosity as the higher
mass stars die out. Note that, because of the underlying power-law
behavior of \Cpar,  on this log scale the curves for different \Cpar\
parameters are offset from each other in proportion to the change in
$\log{\cal C}$.  

\begin{figure}
\plotone{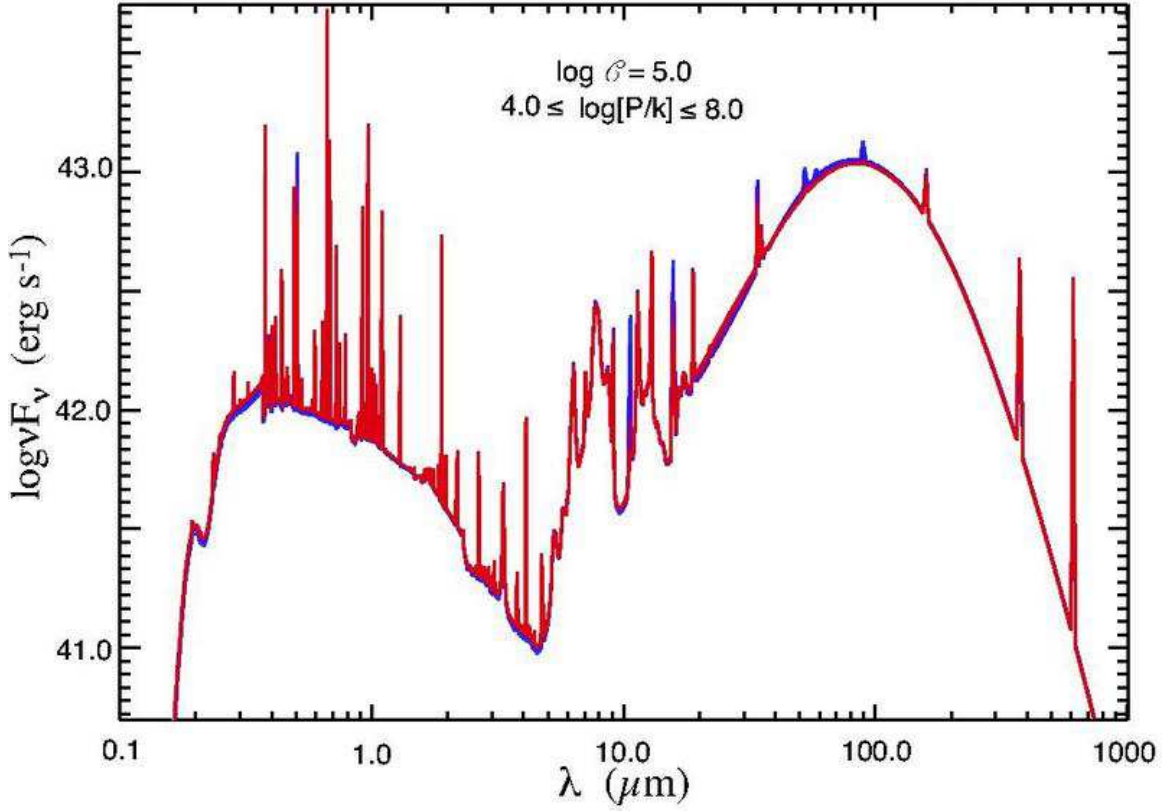}
\caption{Five SEDs with Solar metallicity, $\log \Cpar=5.0$ and
  varying pressure ($\log P/k=4,~5,~6,~7$ and 8). The model SEDs are
  almost indistinguishable apart from their nebula emission features,
  such as the [\neii]12.8 \mum\ emission line. }\label{fig:sameC} 
\end{figure}

In order to demonstrate the constancy of the dust temperature
distribution at constant $\log {\cal C}$,  in figure \ref{fig:sameC}
we show overplotted five SEDs for Solar metallicity photodissociation
regions, which all have the same compactness parameter of $\log{\cal
  C} = 5.0$, but which have different pressures and stellar cluster
masses.  Although these 5 SEDs appear indistinguishable, they are not
exactly the same, because their nebular excitation parameters ($\cal
R$; \emph{see} SED3) are quite different, and consequently their
nebular continuum and emission lines are different.  

Provided that we could independently determine $\cal R$ (from the
nebular spectrum) and  $\cal C$ (from the form of the dust continuum)
then in principle we could solve independently for the mean pressure,
$P_0/k$ and mean cluster mass, ${M_{\rm cl}}$; 

\begin{eqnarray} 
\log \left[ \frac{M_{\rm cl}}{M_{\odot}}\right] = 
\log {\cal C} + \frac{2}{5} \log{\cal R} \\
\log \left[ \frac{P_{0}/k}{\pccm{\rm K}} \right] = 
\log {\cal C} - \frac{3}{5} \log{\cal R}.
\end{eqnarray} 

In practice, the separation of these variables would be assisted by a
direct measurement of the gas pressure. For $P_{0}/k  > 10^6$\pccm K
we can use the ratio of the [\ion{S}{2}] $\lambda 6717,6731$\AA\ lines
for this purpose. 

To show the effect of varying \Cpar\ on the form of the far-IR SED, we
present in 
figure \ref{fig:sameP} six model PDR SEDs having the same
metallicity ($1Z_{\odot}$) and pressure ($P/k=10^5$), with \Cpar\ 
varied by varying the cluster mass.  In the optical and near-IR, the
model SEDs show stellar emission, and the extinction of all six SEDs
is the same, as they pass through the same column depth of dust and
gas. At longer wavelengths, the progression in the dust temperatures
with increasing \Cpar\ is obvious. 

\section{PDR Covering Fraction} \label{sec:fpdr}

\begin{figure}
\plotone{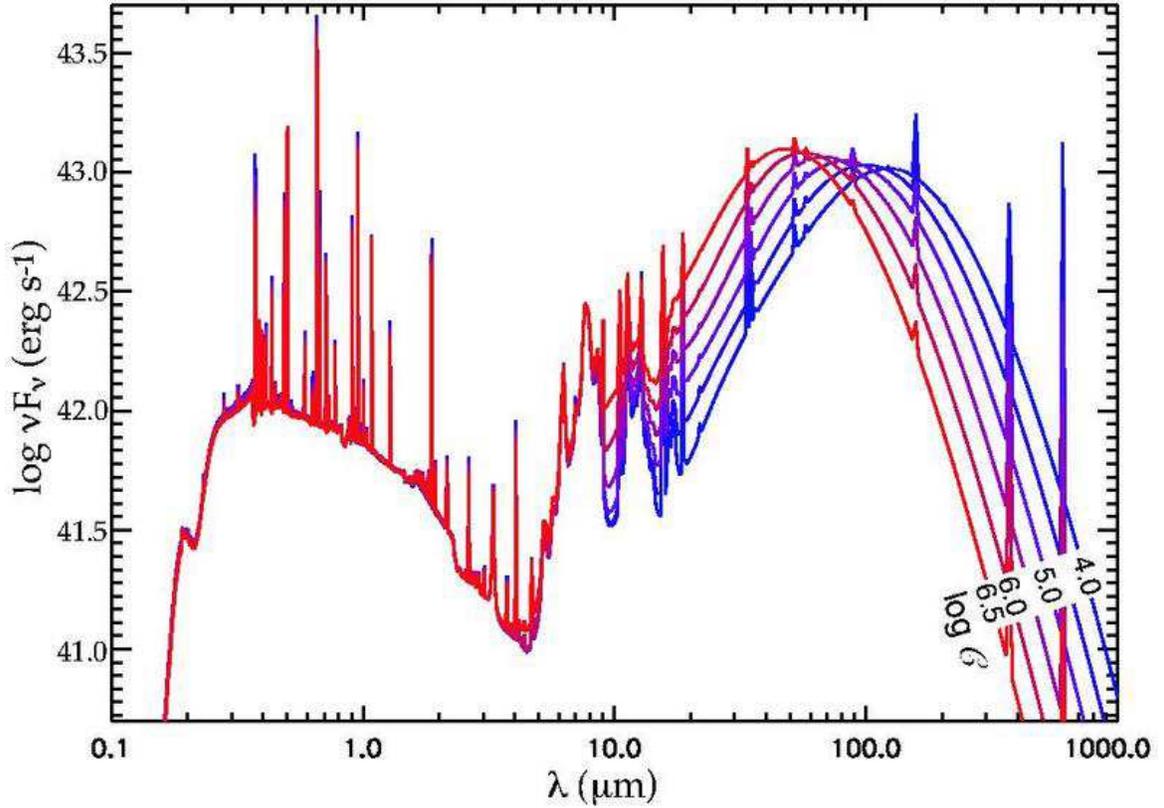}
\caption{Six model SEDs with Solar metallicity, $\log P_{0}/k=5.0$
  and varying compactness. The compactness parameter decreases from
  $\log \Cpar=6.5$ to $\log \Cpar=4.0$ as the far-IR dust emission
  feature moves to longer wavelengths.}\label{fig:sameP}  
\end{figure}

In the previous section, our figures \ref{fig:sameC} and
\ref{fig:sameP} corresponded to a complete covering fraction of
molecular clouds; $f_{\rm PDR}=1$. In this extreme, the molecular gas
and dust surrounding the \HII\ regions act as a dust bolometer,
absorbing essentially all of the stellar UV continuum, and
re-radiating it into the far-IR bump and the PAH features. However, in
the case of isolated \HII\ region complexes in both starburst and in
normal disk galaxies, the placental molecular cloud is quickly cleared
away by the stellar winds, and by photo-evaporation. In older
clusters, the disruption of the cluster by this gas ejection
 will cause the exciting stars to
disperse away from the regions of high extinction, though the
timescale for this may be greater than the \hii\ region lifetime 
\citep{Boily03a, Boily03b}. This process is
cutely referred to as ``infant mortality''. 

Previously in SED1 and SED2, we parameterized this uncovering of the
exciting stars by the introduction of a molecular cloud clearing  or
dissipation timescale, $\tau_{\rm clear}$, where the covering fraction
of molecular cloud PDRs around a stellar cluster is given by;  
\begin{equation}\label{eqn:tauclear}
f_{\rm PDR}=\exp(-t/\tau_{\rm clear}).
\end{equation}
In SED2,  we found that, for Galactic star forming regions at least,
this timescale is quite short, on the order of $1-2$\ Myr. However, this
certainly does not represent all starforming regions, and is probably
far too short for ULIRGs which have an extremely generous sink of
molecular material. In these objects, the \HII\ regions of individual
clusters may merge, but the complex is still surrounded by molecular
gas.  
Thus, the clearing timescale is likely to show a large range, and
will depend on the local environment. In addition, situations like the
commonly observed ``Blister \hii\ regions'', where the star formation
occurs on the edge of a molecular cloud, are not so well represented
by this formalism. 

To deal with this problem, we introduce here the much simpler system
of $f_{\rm PDR}$, which is defined as the time-averaged PDR covering
fraction during the \HII\ region lifetime. Starburst in which the PDRs
entirely surround their \HII\ regions have $f_{\rm PDR}=1$ while
uncovered \HII\ region complexes have $f_{\rm PDR}=0$.  

\begin{figure}
\includegraphics[width=0.9\hsize]{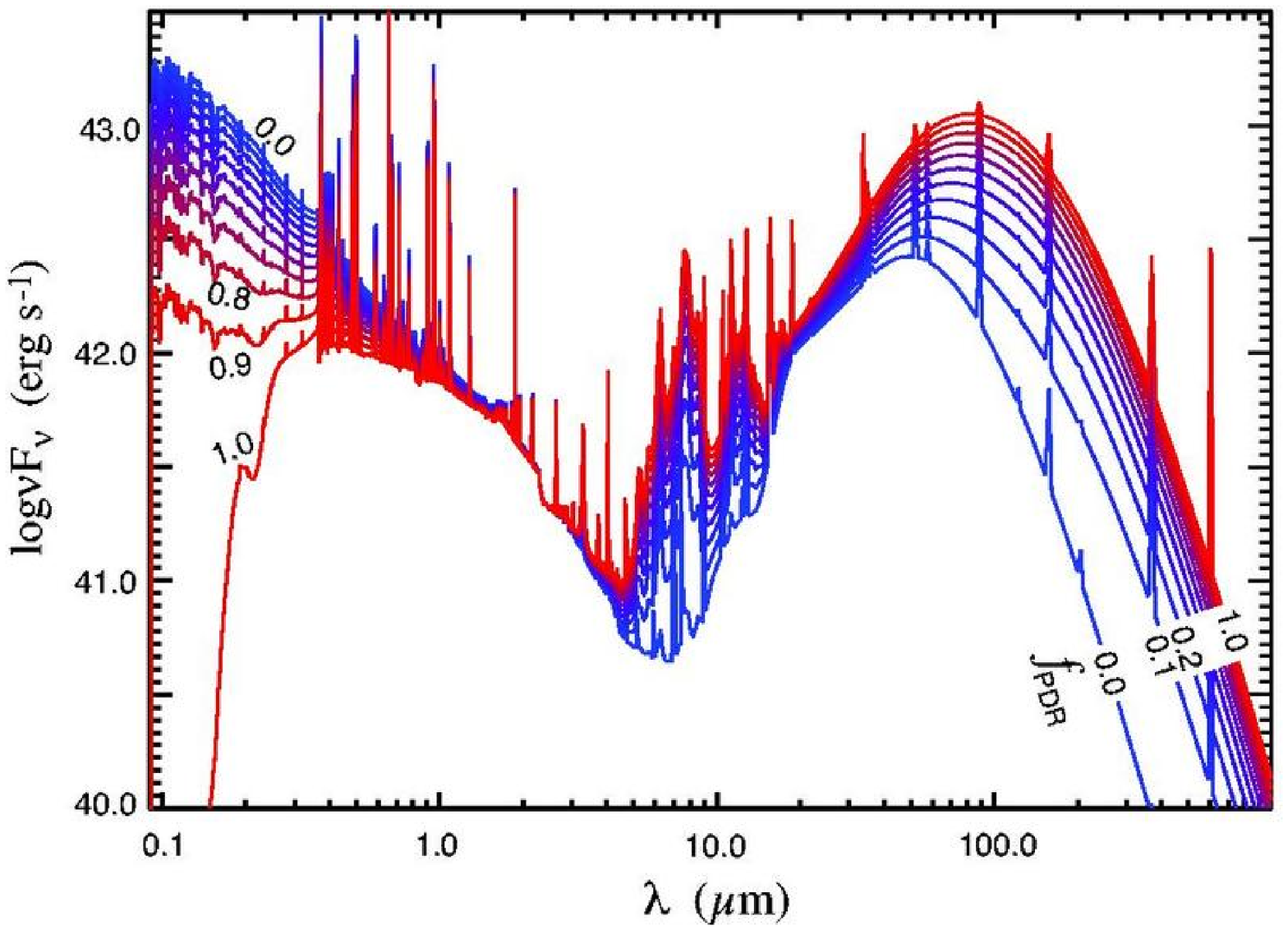}
\includegraphics[width=0.9\hsize]{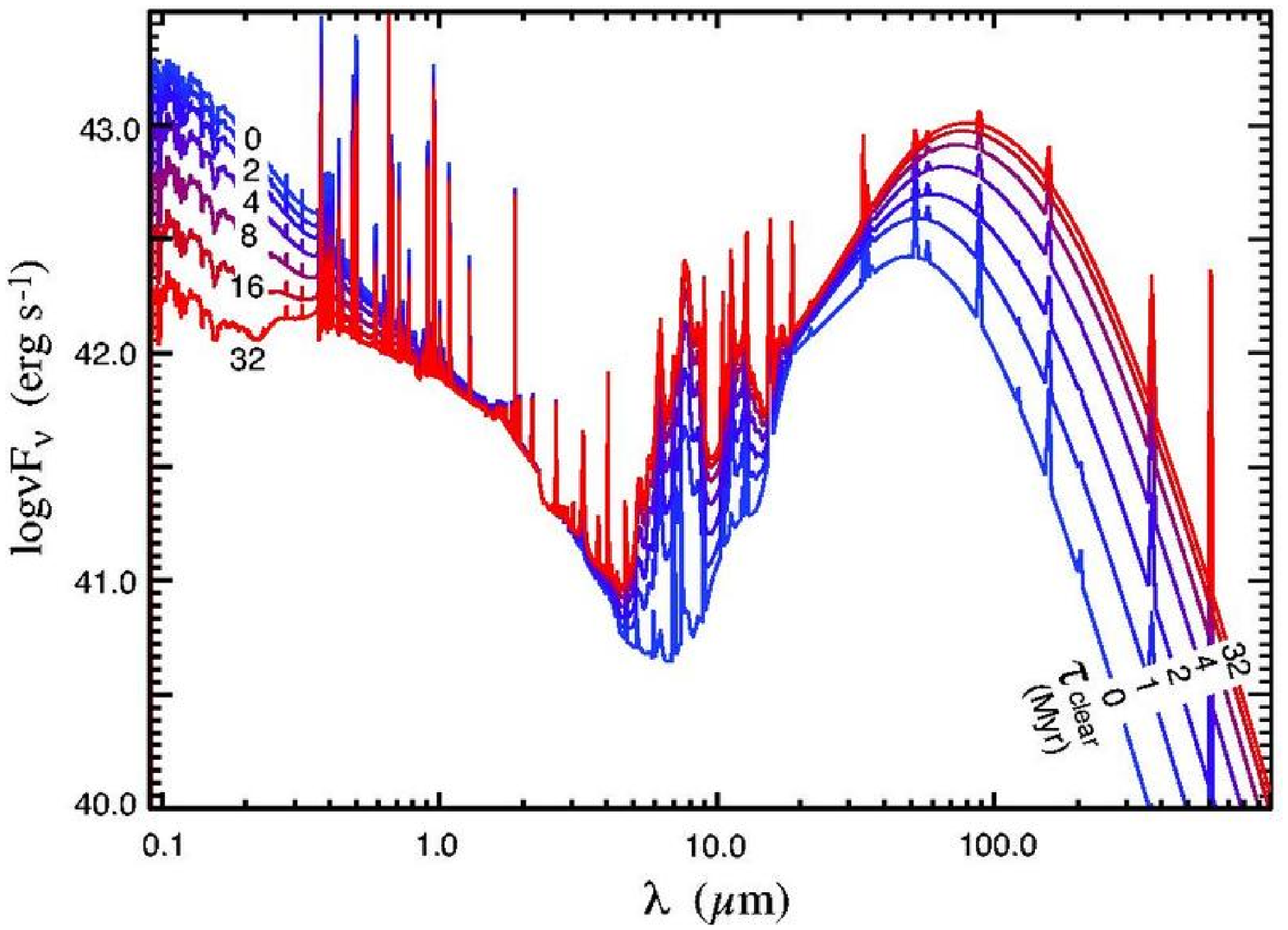}
\caption{SEDs for a Solar metallicity starburst with $\log P/k=5$,
  and $\log \Cpar=5$. The top panel has $f_{\rm PDR}$ ranging from 0.0
  to 1.0 in steps of 0.1. The bottom diagram shows the SEDs using 
the $\tau_{\rm clear}$ parameter introduced in SED1, with $\tau_{\rm 
  clear}=0,~1,~2,~4,~8,~16,$ and 32\ Myr. It is clear that each of
these two formulations of the PDR covering factor problem map to the
other.}\label{fig:tauvsf} 
\end{figure}

In figure \ref{fig:tauvsf} we show two series of SEDs to illustrate
the effect of a changing PDR covering fraction. Both represent a solar
metallicity starburst with a compactness $\Cpar=10^5$ and pressure
$P_{0}/k=10^5$ K\pccm, and are luminosity-weighted integrations scaled
to a star formation rate of 1.0 \Msunpyr.  The upper set of SEDs show
the change in SED using $f_{\rm PDR}$, evolving from a pure \hii\
spectrum ($f_{\rm PDR}=0$) to a pure PDR spectrum ($f_{\rm
  PDR}=1.0$). The \HII\ region-only SED (the same in both series) is
characterized by a strong    
stellar UV continuum, absent or weak PAH features, and hot dust
emission from within the \HII\ region itself. By contrast, the \HII\
region plus PDR models show strongly attenuated 
blue and UV continua ($A_{\rm V}\sim 0.8$ at $f_{\rm
  PDR}=1.0$), strong PAH features and a broad, cool far-IR feature. 

As a comparison, the second plot shows a series in $\tau_{\rm clear}$,
going from $\tau_{\rm clear}=0$\ Myrs (pure \hii\ region) to $\tau_{\rm
  clear}=32$\ Myrs. Clearly, these two formulations of the covering
factor are broadly equivalent.  On close examination it is possible to
see some stronger older star 
features in the $\tau_{\rm clear}$ spectra relative to a matching
$f_{\rm PDR}$ model, but these are small. Concurrent with this is the 
slightly wider IR feature in the $f_{\rm PDR}$ models relative to the
$\tau_{\rm clear}$ 
models due to the stronger presence of dust heated by ``old'' ($\sim
10$\ Myr) star cluster light.  

Note that, as we increase the covering factor, there is an increase in
the mid-IR from around 4\mum\ up to 15\mum\ caused by the increasing
contribution of PAHs. The far-IR dust re-emission feature
progressively increases and broadens as the contribution of cool dust
in the PDR becomes more important.  The contribution of this cool dust
in the PDR can be traced at shorter wavelengths through the steepening 
of the 20-35 \mum\ slope. This
spectral region has few emission lines and there are now available
many \emph{Spitzer} IRS spectra of Starbursts,
e.g. \citet{Brandl06}. Thus the 20-35 \mum\ slope may be a useful
diagnostic of the PDR fraction in starbursts. However, this region is
also affected by the contribution of ultra-compact \hii\ regions (\S
\ref{sec:UCHII}) and by attenuation at high $A_{\rm V}$ (\S \ref{sec:diffuse}).

Note also the insensitivity of the SED to the covering factor in two
regions of the spectrum; the $1-4$ \mum\ region, and at around 20
\mum. The $1-4$ \mum\ emission is dominated by the older stars in the
starburst, and for these the PDR acting by itself is insufficient to
produce a significant dust attenuation. The constancy of flux in the
$\sim 20$\mum\ waveband is somewhat more interesting and is directly
related to the strong correlation between the 24\mum\ \emph{Spitzer}
flux and the SFR \citep{Calzetti05}. The models reveal that almost all
of the warm dust emission arises from the hot dust embedded within the
\hii\ region itself. The cooler dust in the surrounding PDR makes
little contribution to the global flux at these wavelengths. 
This result agrees with the
spatially-resolved observations of nearby galaxies where the 24\mum\
emission peaks in \hii\ regions while the PAH emission is
much more diffuse. In NGC 5194 about 85\% of total galaxy 24\mum\
emission arises within the defined \hii\ regions, while only
$\sim$60\% of the 8\mum\ arises within these regions
\citep{Calzetti05}. This result is also in agreement with the earlier
theoretical calculations of \cite{Popescu00} for the edge-on galaxy
NGC~891. In this galaxy, the star forming disk \HII\ regions have to
be associated with a dominant hotter dust emission component. 

Our models also demonstrate that the relationship between SFR and
24\mum\ emission is not one to one, because the warm dust continuum is
also sensitive to the compactness parameter,  \Cpar\ , as is
demonstrated in figure \ref{fig:sameP}. This finding also agrees with
the observations, since variations of two or three in the ratio of the
24\mum\ emission to the SFR have been observed between galaxies
\citep{Dale01,Calzetti05}.  

\section{Column Depth in the Photodissociation Region}

\begin{figure}
\plotone{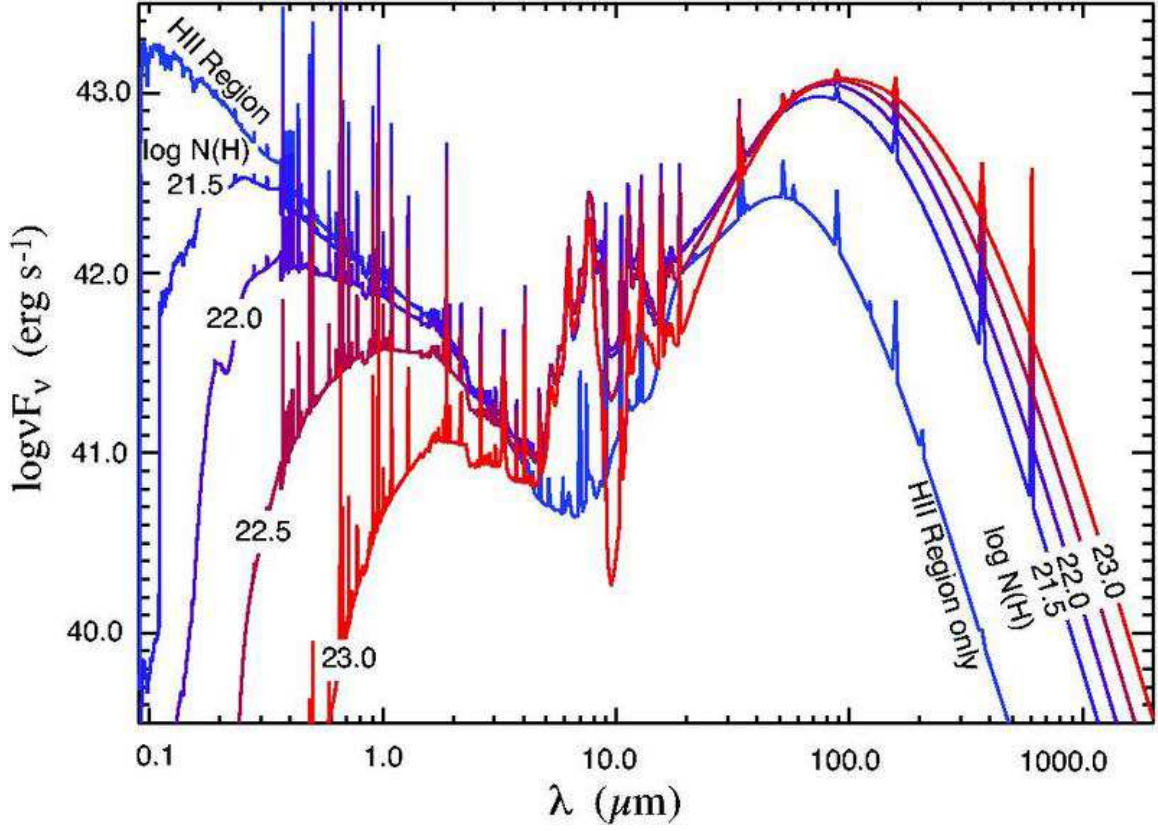}
\caption{Spectra showing the effect of varying the PDR column depth on
a solar metallicity starburst with $\log
\Cpar=5$ and log$P/k=5$. Shown are a PDR free (\hii\ region) SED, and
PDRs with \hi\ columns of $\log N({\rm HI})=21.5$, 22.0 (standard
model), 22.5, and 23.0, as labelled.}\label{fig:column}
\end{figure}

As stated in section \ref{sec:models}, we use a hydrogen column density of
$\log N({\rm HI})=22.0 $ (cm$^{-2}$) to define the extent of the
photodissociation region. While this value is typical of molecular
clouds in our own galaxy, it is quite likely that those in starburst
regions cover a broader range in column depths. In figure
\ref{fig:column} we show the effect of varying the column density in
the model. For this exercise, we use our fiducial starburst model with
solar metallicity,  compactness parameter of $\Cpar=10^5$, ISM
pressure of $P_{0}/k=10^5$K \pccm and test a range of PDR column
depths; $\log N({\rm HI})\sim 0$ (corresponding to the \hii\ region
spectrum), 21.5, 22.0 (our standard PDR model spectrum), 22.5, and
23.0.  

Note that, as more of the UV and visible photons are absorbed, the
total flux in the far-IR bump becomes correspondingly greater, and the
feature also becomes wider as the contribution of the colder dust
heated by softer photons deep within the PDR becomes more
important. Note also that the contribution of the PAHs to the SED is
apparently almost complete by $\log N({\rm HI})=21.5$. 
 
In many respects, the effect of increasing the column depth of the PDR
is similar to that obtained by applying a diffuse dusty screen (see
the following section). As we increase in column depth the optical
depth increases (relative to the \hii\ model). The effective model
$A_{\rm V}$ for each model is 0.2, 0.8, 2.6, and 8.4 as $N({\rm HI})$
increases from 0 to 10$^{23}$ \pscm\ . At $N({\rm HI})=10^{23}$\pscm\
the IR starts to become optically thick, with a corresponding
steepening in the  20\mum\ to 30\mum\ slope. The increase in the
silicate feature depth, $\tau_{9.7 \mu{\rm m}}$, becomes obvious
somewhat earlier,  at $N({\rm HI})=10^{22}$\pscm. Note that
differences in the dust composition and dust geometry ensure that the
attenuation law of the diffuse dust (shown in fig. \ref{fig:extcurve})
is slightly different to that experienced in the PDR, with the
silicate grains more prominent in the PDR (as seen by the stronger
silicate absorption feature and flatter extinction).  

The main difference in our model between the PDR region and a  dusty
screen is the
inclusion of the self-consistent dust emission. By $\log
N({\rm HI})=22$ most of the UV-optical flux has been absorbed
and re-emitted as the strong FIR feature. At larger depths
the average temperature of the dust is quite cold ($\sim
10-20$K) and emits only at long ($>100\mum$) wavelengths. 
However in starburst regions such temperatures may not be reached, as
 it is likely that neighboring clusters, as well
as the underlying diffuse population, would prevent the dust from
reaching such temperatures. Only in the largest molecular clouds, or
the most distant dust, would such temperatures be reached. It is for
this reason that we limit our photodissociation regions to $\log
N({\rm HI})=22.0$ and use the diffuse population to represent this
cool dust (\S \ref{sec:diffemiss}).

\section{Other Components of the SED}

\subsection{Ultra-Compact \HII\ Regions}\label{sec:UCHII}

The modeling presented thus far assumes that the stars in
stellar clusters inflate a common \HII\ region, and that the cluster
can be treated as a single, centrally concentrated source of radiation
and mechanical luminosity. However, in the earliest stages of the
cluster lifetime ($\lapprox 10^6$ years after formation) the
individual stars composing the clusters are likely to be still buried
in their separate birth clouds. In addition, we know that the massive
stars start burning hydrogen even before they reach the main sequence
and while they are still accreting matter from their parental
molecular cloud \citep{Bernasconi96}. During this phase, the cluster
acts as an ensemble of Ultra-Compact \HII\ regions (UCHII), each
trapped at sub-parsec scales around their individual massive parent
stars  \citep[see][ for a detailed review of UCHII
  regions]{Churchwell02}. The period of time in which the winds of
cluster stars cannot operate collectively may occupy a significant
fraction of a massive star's lifetime \citep[][ and references
within]{Rigby04}. During this 
UCHII region phase the optical emission will be totally obscured, with
$A_{\rm v} \sim 50$mag. As the UCHII region is so compact, it
also displays a hot IR emission \citep{Peeters02b}. This compactness
also ensures that the dust is very successful in competing against the
gas for the ionizing photons \citep{Dopita03}, which makes the region
still more compact and ensures that radiation pressure effects in the
ionized plasma are important. Thus, UCHII regions are qualitatively
different from normal cluster-driven \HII\ regions. 

The technique of modelling the SEDs of individual compact and
ultra-compact \HII\ regions was fully described in
\cite{Dopita05b}. Briefly, these use the \emph{TLUSTY} models
(available at  http://tlusty.gsfc.nasa.gov) which were interpolated
and re-binned to the energy bins used in our code, \mapiii. On the
basis of the results presented by \cite{Morisset04}, we can expect the
SEDs we derive will be very similar to those using either the
\emph{WM-Basic} or the \emph{CMFGEN} models. Unlike the \emph{TLUSTY}
atmospheric models, these latter two are fully dynamical atmospheric
models. The \emph{TLUSTY} models used cover three abundances, 0.5, 1.0
and $2.0Z_{\odot}$. Here the definition of solar abundance is
effectively the same as used in the Starburst99 models. 

We made \mapiii\ photo-ionization models for the structure of
radiation-pressure dominated dusty \HII\ regions and their surrounding
photo-dissociation regions. The models are started close to the
central star (1\% of the computed (dust-free) Str\"omgren radius) so
that the ionized gas effectively fills the Str\"omgren  sphere of
these models. We use the same dust model as for the cluster SED
computations. The models that we used here have fixed external
pressure $\log P/k = 9.0 $ K cm$^{-3}$, and are terminated at a \HI\ column
density of $\log N({\rm HI}) = 21.5$. The mass range of the central
stars is $16.7-106.9 M_{\odot}$. This corresponds to a stellar
effective temperature range of $32500 \leq T_{\rm eff} \leq 52500$K,
which is covered in bins each 2500K wide. The computed SEDs for
individual UCHII regions correspond to the zero age main sequence
(ZAMS) of the central stars. 

The starburst galaxy compact \HII\ region SED is constructed by
co-adding the SEDs computed for each stellar mass bin, luminosity
weighted to that corresponding to a Kroupa Initial Mass Function (over
the effective mass range $15-120 M_{\odot}$), and scaled to represent
a massive star formation rate of $ 1.0 \msolar$ yr$^{-1}$ continued
over a period $0.0-1.0$\ Myr.  

\begin{figure}
\plotone{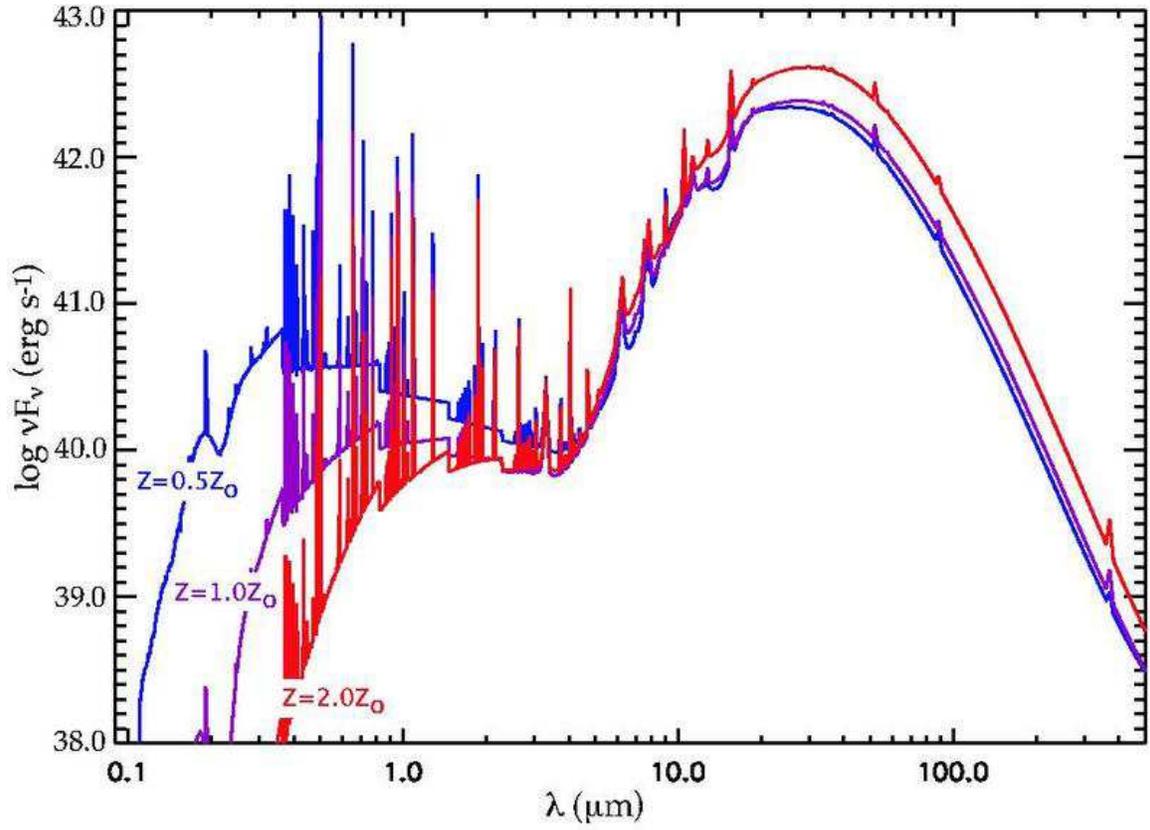}
\caption{The UCHII region templates for 0.5, 1.0 and $ 2.0Z_{\odot}
  $. These correspond to a Kroupa IMF and a star formation of  $1.0
  \msolar$ yr$^{-1}$ continued over 1.0\ Myr. The small changes in the
  apparent normalization of these spectra is due to the intrinsic
  change of stellar luminosity with
  abundance. }\label{fig:UCHII_SEDs} 
\end{figure}

The resulting SEDs for each metallicity are shown in figure
\ref{fig:UCHII_SEDs}.  The small changes in the apparent normalization
of these spectra is due to the change of the stellar luminosity with
abundance. Note that all the three spectra are very similar, with very
weak PAH features, a hot dust far-IR continuum, similar line spectra
and heavy (but abundance-dependent) obscuration at shorter
wavelengths. 

Because the spectra are quite similar, it is clear that any one of
them could be used as a UCHII template across the full metallicity
range. However, for closest consistency,  we recommend the use of the
$2.0Z_{\odot}$ model with the  $2.0Z_{\odot}$ Starburst99
cluster templates, the  $1.0Z_{\odot}$ model with the  $1.0Z_{\odot}$
Starburst99 cluster models and the  $0.5Z_{\odot}$ model with
the  0.05, 0.2 and 0.4$Z_{\odot}$ Starburst99 cluster models. 

\begin{figure}
\plotone{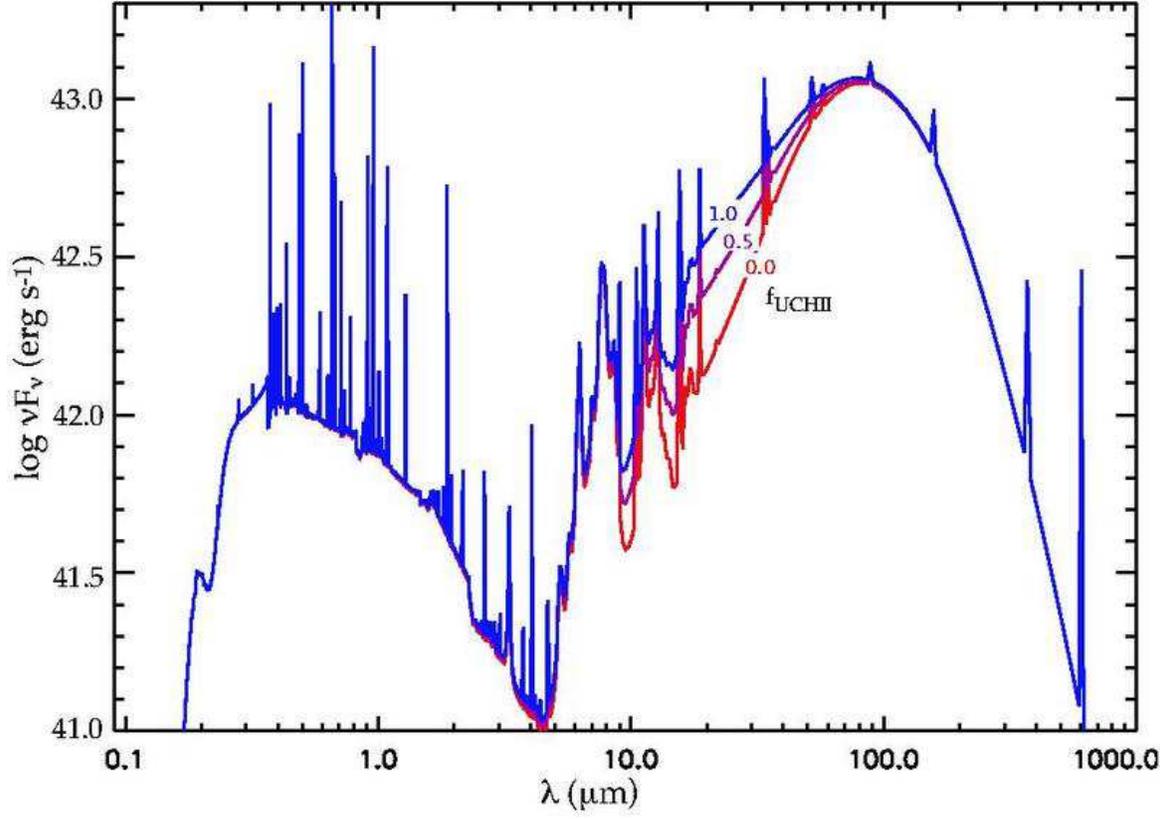}
\caption{Three SEDs of a solar metallicity starburst with $\log \Cpar=5$,
  log$P/k=5$ and $f_{\rm PDR}=1.0$. Shown is the effect of including the
  $1Z_{\odot}$ UCHII template with $f_{\rm UCHII}=0.0$, $0.5$,
  and $1.0$, with the mid-IR continuum increasing with increasing
  $f_{\rm UCHII}$.}\label{fig:addUCHII}  
\end{figure}

As an example, in figure \ref{fig:addUCHII} we demonstrate how the
inclusion of UCHII emission alters the starburst SED. We parameterize
the addition of this emission to the SED by $f_{\rm UCHII}$, 
the scale of the UCHII
contribution. A $f_{\rm UCHII}=1.0$  implies
that 50\% of the massive stars younger than 1.0\ Myr are surrounded by
ultra-compact \hii\ regions. As the UCHII regions emit predominantly
in the mid-IR, this parameter affects both the mid-IR
slope and PAH equivalent widths, as seen in figure \ref{fig:addUCHII}.
The emission lines are also affected by the inclusion of the young
\hii\ regions, with the contribution greater in the mid-IR than the
optical due to the high optical depth of the UCHII regions.

\subsection{The Older Stellar Contribution}\label{sec:oldstar}

The discussion so far has concerned itself only with the stars younger
than 10\ Myr. However, a typical starburst will continue for a dynamical
timescale of a galaxy, typically $\sim10^8$~yr. Over this time period,
most of the the young star clusters will have dispersed into the
field, and away from the active star forming regions
\citep{Whitmore07, Fall05}. As a consequence, the older ($>10$\ Myrs)
stellar population may dominate the optical and UV parts of the
SED, since it suffers much less extinction than the starburst itself,
having long escaped its original molecular birth clouds
\citep{Charlot00}.  
However, it is in the optical-UV range that most of the work has been
done in constraining the star formation history (SFH) of galaxies,
using features such as the 4000\AA\ break due to hydrogen and stellar
absorption lines like the Lick indices to constrain the age and
metallicity of the dominant stellar population, or even the evolution
of star formation and metallicity \citep[eg][]{Panter07}. 

\begin{figure}
\plotone{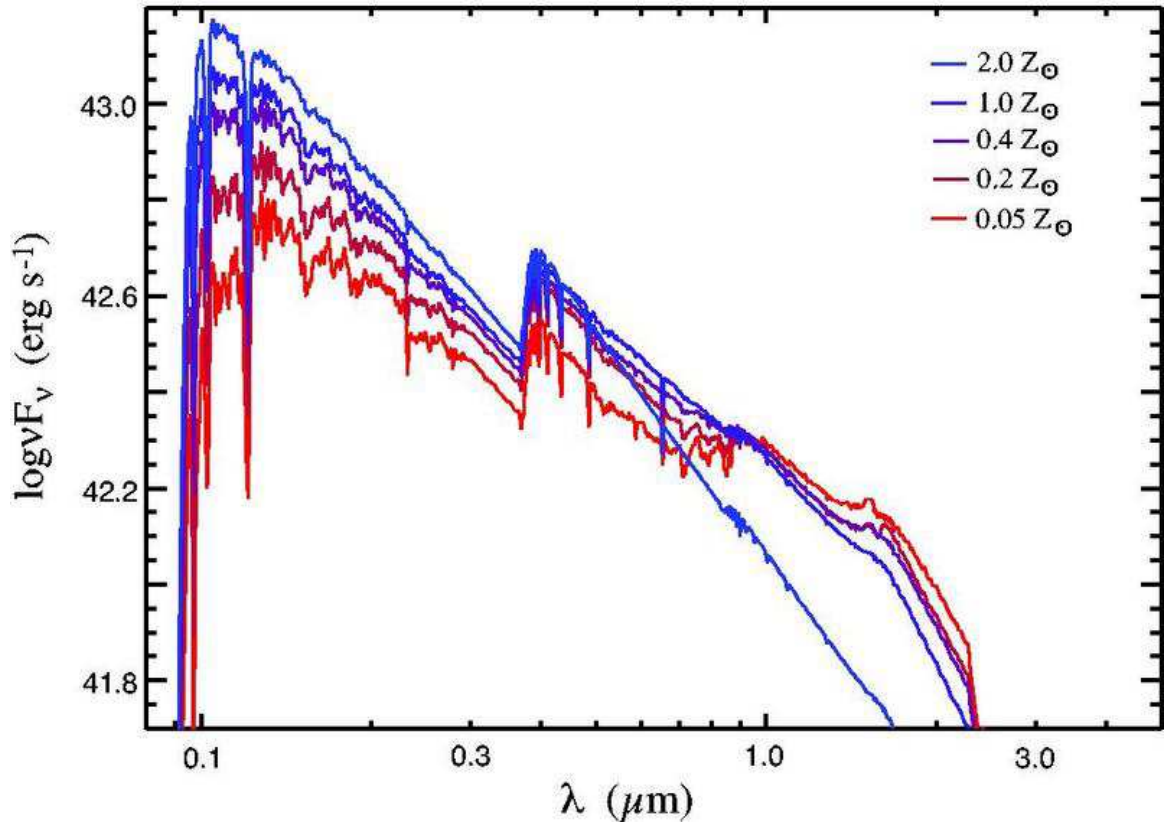}
\caption{Spectral Energy Distributions of our ``old'' stellar
populations for the five starburst metallicities (as labelled) and
scaled to a continuous SFR of 1\Msunpyr. }\label{fig:oldstar} 
\end{figure}

In order to represent this older stellar contribution in our starburst
models we use a luminosity weighted sum of the 10-100\ Myr starburst
spectra from Starburst99, having the same parameters as the
models used to generate the \hii\ spectra. We generate the ``old''
stellar spectra for each metallicity, and assume that the metallicity
of the starburst the old population are the same. The resulting
spectra for each metallicity are shown in figure \ref{fig:oldstar},
scaled to a continuous star formation rate of 1.0\Msunpyr.  

\begin{figure}
\plotone{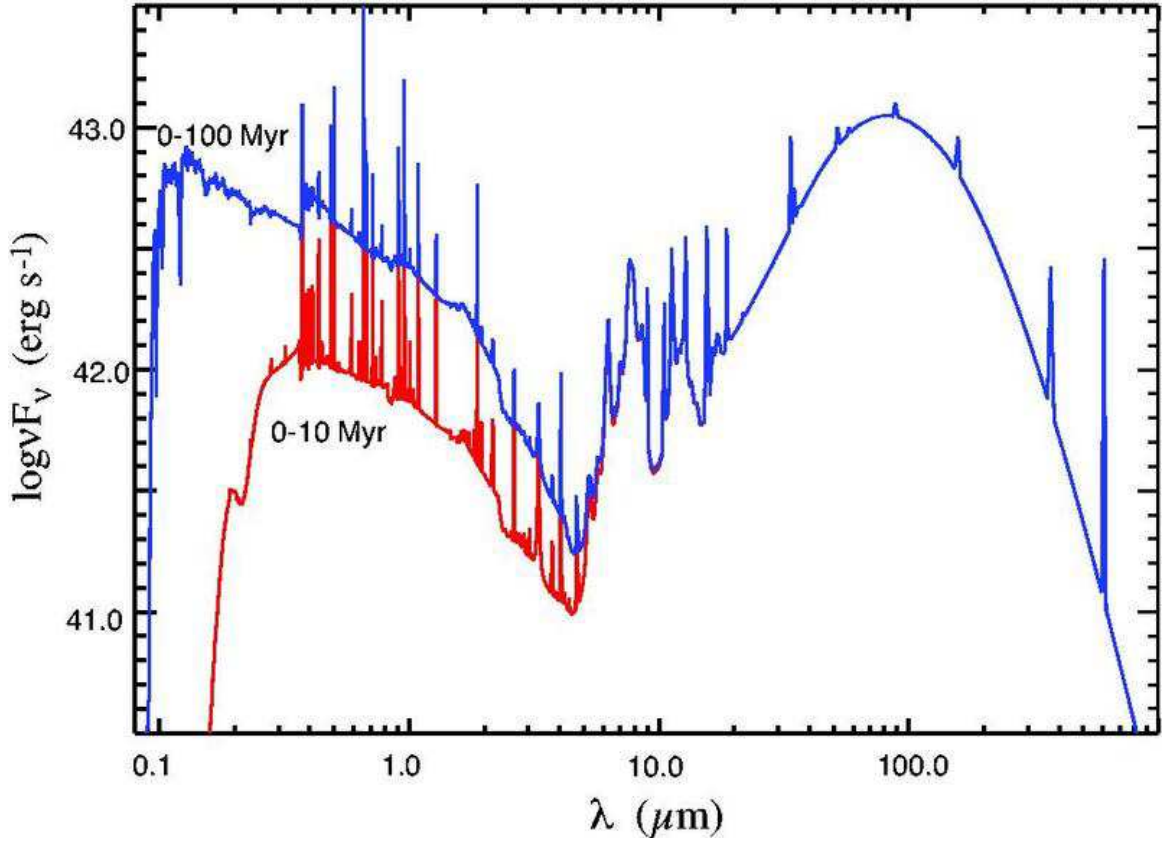}
\caption{SED of a starburst with $1Z_{\odot}$, $\log \Cpar=5$,
  log$P/k=5$ and $f_{\rm PDR}=1.0$ (lower SED) and SED of the same
  starburst with our $1Z_{\odot}$ ``old'' stellar SED added (upper
  SED). We assume a constant SFR of 1\Msunpyr\ for the computation of
  both curves.}\label{fig:hii+old}   
\end{figure}

\begin{figure}
\plotone{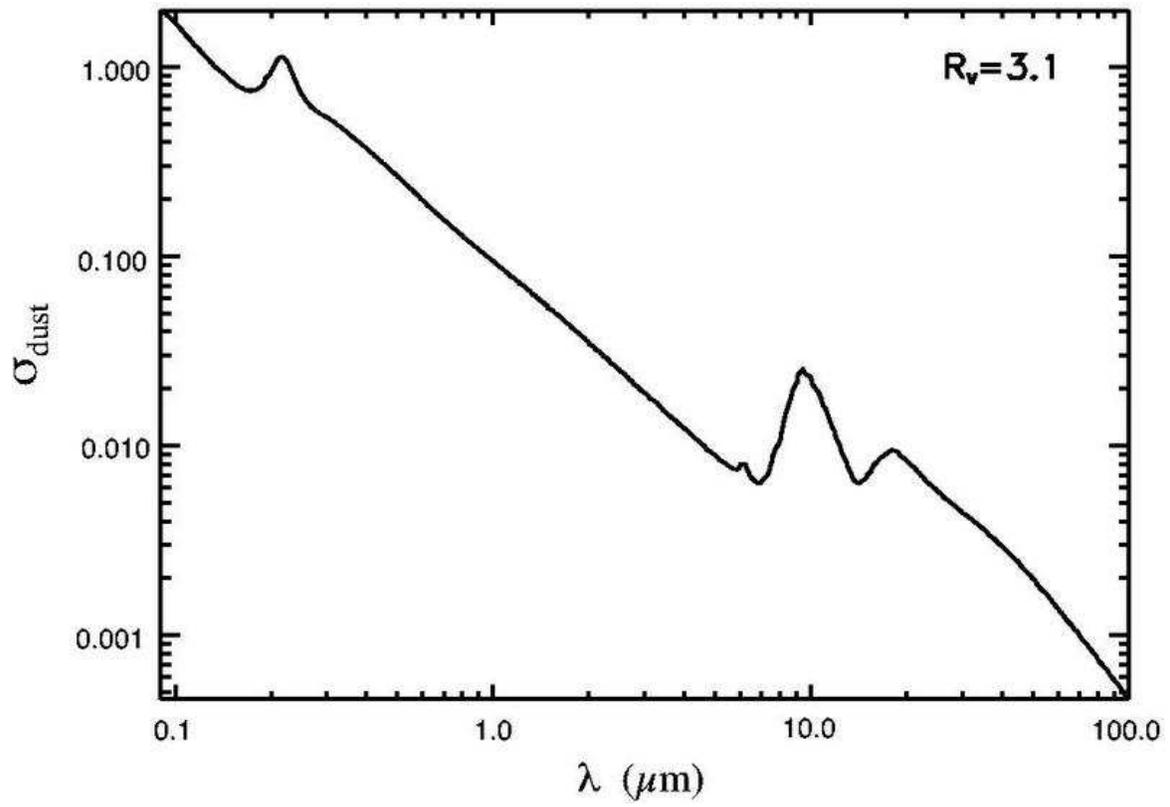}
\caption{Attenuation curve from \citet{Fischera05} with an $R_{\rm
    V}=A_{\rm V}/E_{\rm B-V}=3.1$. Note that the 2175 \AA\ carbon ring
  $\pi -$orbital resonance diffuse absorption feature is weak relative
  to the extinction curves measured in the Milky Way. Note also the
  silicate features at 9.7 and 18 \mum.} \label{fig:extcurve} 
\end{figure}

The inferred ratio of the starburst fraction in the $< 10$\ Myr
population to the fraction in the age range 10-100\ Myr  can provide
information about the progress of the starburst - whether the
starburst activity is accelerating or decelerating in recent
time. This fraction may be constrained by comparing the flux at some
wavelength in the far-IR bump with the stellar continuum flux at any
wavelength shorter than about 5\mum. In figure \ref{fig:hii+old} we
demonstrate this sensitivity. Here we have added this old stellar
contribution to a solar metallicity starburst with $\log \Cpar=5$ and
log $P/k=5$. We have normalized the SED, assuming a continuous and
constant star 
formation rate of 1 \Msunpyr\ up to the maximum starburst age of
$10^8$yr. To emphasize the effect of the older population we have
taken a totally obscured $<10$\ Myr starburst ($f_{\rm PDR}=1$), as
shown in the lower curve, and added a completely unobscured
contribution from the older stars to provide the upper curve.  

These models show how it is possible to produce a systematic
difference between the dust attenuation in the UV and the dust
attenuation of the \HII\ regions as measured by the Balmer
Decrement. Such a systematic difference has long be known to exist
from observations of starbursts \citep{Calzetti94, Calzetti01}, in the
sense that the Balmer Decrements indicate greater attenuation than the
UV and visible stellar continuum.  

Finally, the ratio of the two stellar components, young/old, is
related to the $b$ parameter recently used by \citet{Kong04}, which is
the ratio of the current versus past-averaged star formation
rate. This parameter was introduced to help understand the
relationship between the UV and IR, through concentrating in
particular on the correlation of the UV slope and IR excess
\citep{Meurer99}. In our case we parameterize the contribution of the
old stellar population  through the parameter $f_{\rm old}$.
Figure \ref{fig:hii+old} presents a case of $f_{\rm old}=1$,
with continuous star formation, and $f_{\rm old}$ greater and lower
than one represents cases where the past average star formation
history is greater and less than the current star formation
respectively. 

\subsection{Diffuse Dust Attenuation}\label{sec:diffuse}

The simple \hii\ region plus photodissociation region models presented
in the previous sections provide a maximum attenuation of $A_{\rm
  V}\sim0.8$ for a solar metallicity starburst and $A_{\rm V}\sim2.5$
for 
$2Z_{\odot}$. This is lower than observed in many ULIRGs
\citep{Farrah07}. In order to properly account for the total
attenuation, we need to include the attenuation produced by a by a
foreground dusty screen associated with gas in the starburst host
galaxy, but not necessarily partaking in the starburst, to account for
these heavily obscured starbursts.  This foreground screen will also
attenuate the older stars associated with the starburst (discussed in
the previous section).

The properties of such a turbulent foreground attenuating screen were
discussed in a series of papers by Fischera \&
Dopita \citep{Fischera03, Fischera04, Fischera05}. They showed that
turbulence which produces a log:normal distribution in local density
will also, to a high level of approximation, produce a log:normal
distribution in column density. They also showed that the resulting
attenuation curve  is unlike that of a normal extinction law, showing
lower attenuation in the UV and larger attenuation in the IR, due to
the spatially-varying extinction across the face of the dust-obscured
object. Here we adopt the theoretical attenuation curve computed in
those papers, which is shown in fig.~\ref{fig:extcurve} and which
provides a close approximation to the empirically-derived Calzetti
extinction law for starburst galaxies \citep{Calzetti01}. This curve
does not allow for the possible destruction of PAHs in the diffuse
medium, and the computed 2175 \AA\ absorption feature may be rather
too strong to be applicable to starburst environments.  

\begin{figure}
\plotone{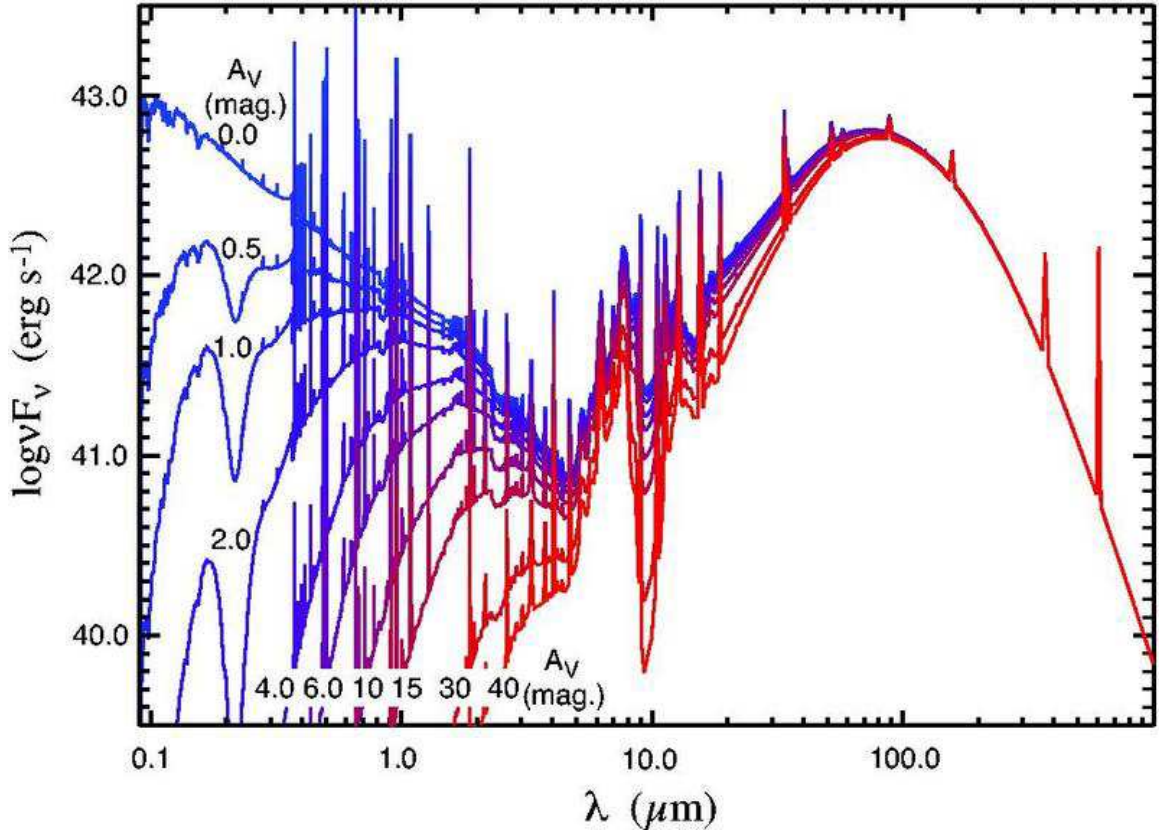}
\caption{The SED of a model starburst with solar metallicity, $\log
\Cpar=5$, log$P/k=5$, and $f_{\rm PDR}=0.5$ showing the effects of
increasing attenuation by diffuse dust. Applying the attenuation curve
of figure \ref{fig:extcurve}, we show $A_{\rm V}$ of 0.0, 0.5, 1.0,
2.0, 4.0, 6.0, 10.0, 15.0, 30.0 and 40.0 as labelled and seen by the
decreasing UV and optical flux. Note that the 
diffuse dust emission associated with the attenuation has not been
included.}\label{fig:atten}
\end{figure}

In figure \ref{fig:atten} we show the effects of this dusty screen on
a starburst with solar metallicity, compactness parameter of
$\Cpar=10^5$, ISM pressure of $P_{0}/k=10^5$K \pccm, and covering 
fraction of $f_{\rm PDR}=0.5$. We show both low and high attenuation,
with 10 SEDs plotted in figure \ref{fig:atten} with $A_{\rm V}$ of
0.0, 0.5, 1.0, 2.0, 4.0, 6.0, 10.0, 15.0, 30.0 and 40.0, with a clear 
depletion of the ultraviolet and optical flux as the $A_{\rm V}$
increases. To emphasize the effects of the attenuation on the IR
emission we do not include the diffuse cool-dust emission that would 
be expected with such attenuation. We note that beyond $\sim 60$\mum\
there would be a contribution due to thermal emission from this cool
dust, and care should be taken with any interpretations made beyond
this wavelength. The effect of this emission is discussed in the
following subsection. 

The attenuation and reddening of the underlying starlight at the
optical and UV wavelengths is very evident. However, it is only at
$A_{\rm V}\gapprox5$ that absorption in the 9.7 \mum\ and 18 \mum\
silicate features becomes apparent. With $A_{\rm
  V}/\tau_{9.7\mum}\sim16.6$ \citep{Rieke85} 
this feature is weaker than the optical opacity, but it is observed to
be visible or very strong in a large number of starburst galaxies,
proving that a good deal of the starburst activity will be totally
obscured in the visible thanks to large column densities in the
surrounding molecular gas. Note also that at the highest $A_{\rm V}$
even the 20-35\mum\ slope steepens to the ``reddening'' effect of the
dust attenuation.

The $A_{\rm V} > 20$mag.~dusty screen needed to provide the observed
depth of the silicate absorption troughs seen in some ULIRGs has
little effect on the global energy balance of the starburst. By an
$A_{\rm V}$ of 1.0, most of the FUV-optical radiation from stars, almost 80\%,
has been absorbed (or scattered) by dust. The
decline in the luminous flux through the PDR, accompanied with the
softening of the radiation field naturally produces the dust
temperature gradient  which \citet{Levenson07} believe is required in
the obscuring material.  

\subsection{Diffuse Dust Emission \& Scattering}\label{sec:diffemiss}
A complete model of a starburst environment should not neglect the
contribution to the far-IR emission by diffuse cool dust. Some of this
may well be the same dust that produces the foreground screen
absorption. The optical photons that are absorbed by the diffuse
galactic dust at high $A_{\rm V}$ are still capable of heating the
dust in the diffuse ISM. In addition, the star forming regions may not
be fully cocooned by its surrounding PDR cloud, so that some fraction
of the cluster UV radiation may escape and heat this diffuse
dust. This cool galactic dust will mostly contribute to the SED in the
$>100$\mum\ to sub-mm region. A portion of this radiation may also be
scattered into our sight-line, which mostly affects the far-UV SED. In
our modeling we have not included this component due to our limited
geometry.   

The exact temperature distribution of the grains will also depend upon
the geometry of the starburst, 
pre-existing stellar population, and dust within the starbursting
galaxy.  Due to the limited geometry of our simple models, we are
unable to model such distributions. Instead we follow the work of
\citet{Dale01}, and calculate the diffuse cool dust emission in terms
of the average  interstellar radiation density. This allows us to model
both distributed starbursts where the average radiation field is high,
and nuclear starbursts, where the rest of the galactic scale dust is
only weakly heated. 

To represent the average radiation field we have used the
luminosity-weighted average of a Starburst99 cluster from
$10-100$\ Myr discussed in the section \ref{sec:oldstar}. While this
is 
younger than the average age of the pre-existing stars in a starburst
host galaxy, these stars are nonetheless likely to provide the
dominant dust-heating radiation field.     

\begin{figure}
\plotone{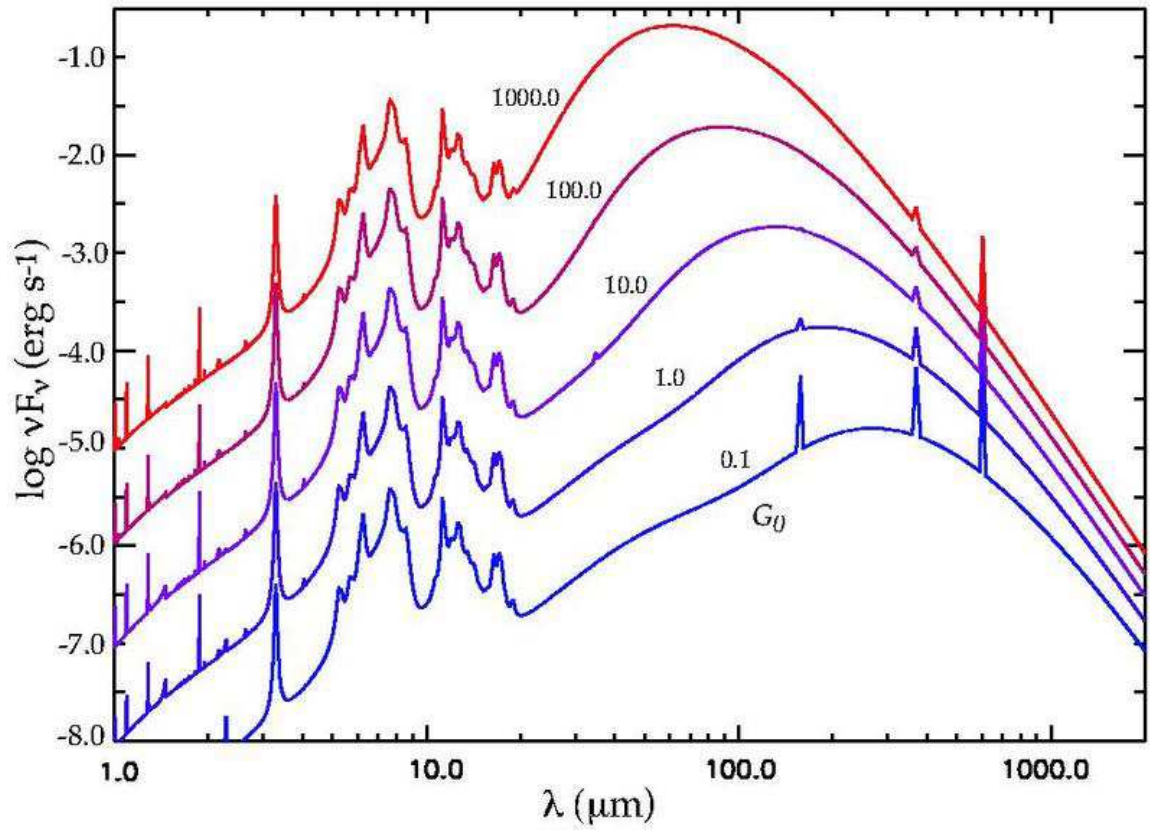}
\caption{Cool dust emission for a Habing radiation field density,
  $G_{0}$, of 0.1, 1.0,  10.0, 100.0 times the Local Interstellar
  Radiation Field Density of  \citet{Habing68}. The strength of the
  radiation field increases from bottom to top.}\label{fig:cool}  
\end{figure}

We then calculate the cool dust emission using the \mapiii\ code,
assuming the same dust properties as before and a column depth of
$\log N(H)=22.5$ (\pscm). For the heating flux we scale the radiation field in
terms of the local \citet{Habing68} interstellar radiation field
(ISRF: FUV$\sim 1.6\times10^{-3}$\flux). We set the radiation field
from 0.1 to 1000.0 times this value, in steps of 0.3 dex. The
resulting  IR emission is shown in figure \ref{fig:cool}. As expected,
low ISRF leads to very cold dust, and the high ISRF has dust
temperatures similar to those encountered in our PDRs. 

The actual contribution of this diffuse dust emission component to the
global starburst SED is not constrained by these models. This would
require a more sophisticated geometrical model of the starburst, its
outflows, and any more extended disk or tidal structure around the
starburst core.  

\begin{figure}
\plotone{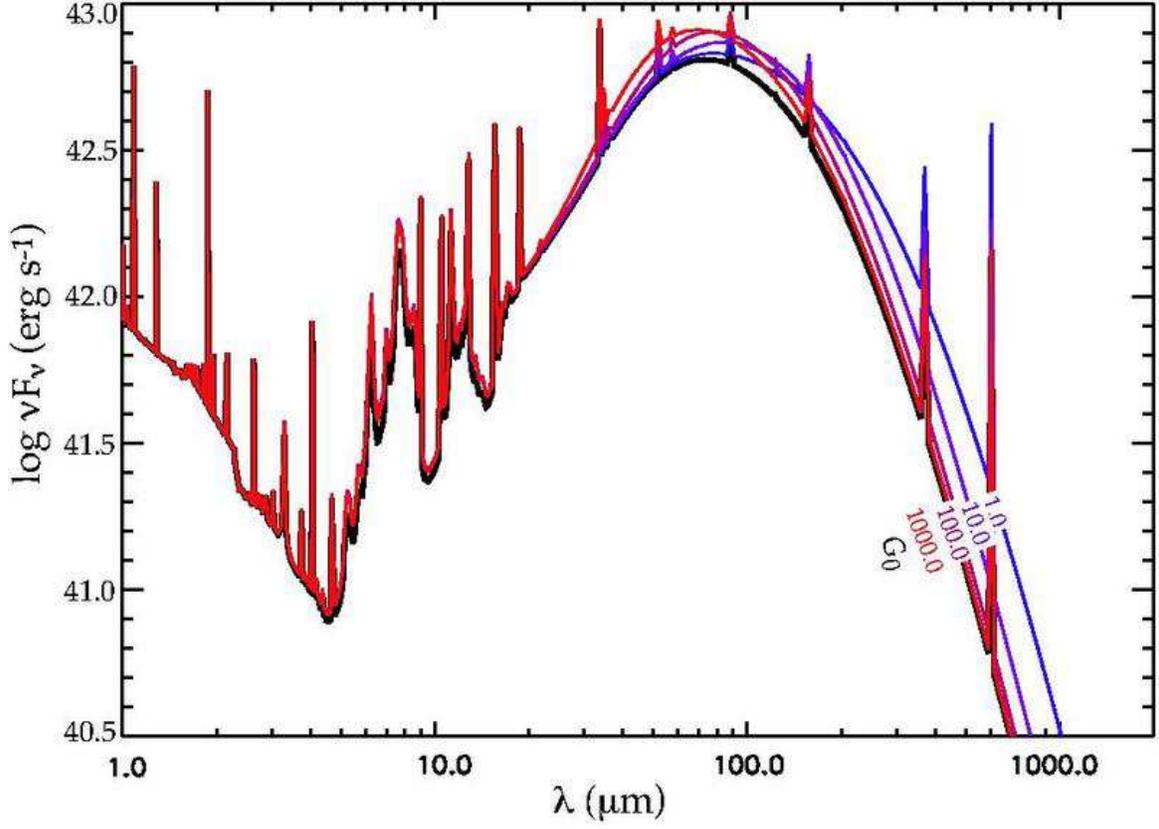}
\caption{Five SEDs demonstrating the effect of adding diffuse dust
  emission at 10\% of the 
Starburst luminosity to our fiducial model Starburst ($1Z_{\odot}$,
$\log \Cpar=5$, log$P/k=5$, and $f_{\rm PDR}=0.5$). The SEDs show the
original SED (thick line) and those with an added diffuse dust 
continuum heated by a young stellar continuum 100.0, 10.0, 1.0 and 0.1
times that of the local ISRF.}\label{fig:adddiff} 
\end{figure}

In Figure \ref{fig:adddiff} we show the effect of adding this diffuse
dust emission component with a total intensity of 10\%  of our
fiducial model Starburst with solar metallicity. For clarity we add
only four of 
the diffuse emission models, heated by a Habing radiation field, $G_{0}$
of 1000.0, 100.0, 10.0, and 1.0 times the local interstellar radiation
field (ISRF$_{\rm local}$), respectively. The attenuation associated
with the diffuse dust emission is not included here.  
 
The model with 1000.0 times ISRF$_{\rm local}$ has hotter dust than the
log \Cpar=5 PDR and is therefore likely to be unphysical. The
100.0 times ISRF$_{\rm local}$ model has a similar dust temperature as 
the PDRs, and so the diffuse field serves only to increase the
total flux of the IR. This probably represents the extreme
case for a starbursting galaxy, and may lead to the  narrow and strong
far-IR features seen in some ULIRGs.

The cooler diffuse emission models, with $G_{0} <$ 10.0 times 
ISRF$_{\rm local}$, 
 are both applicable to less energetic
starburst galaxies.  In such cases, the diffuse emission acts to
broaden the IR feature, as well as shift the peak to longer
wavelengths, while leaving the shorter wavelengths ($<60$\mum)
relatively unaffected. 

\section{Starburst Metallicity}\label{sec:metal}

The metallicity of a starburst affects the SED in several ways;
through the intrinsic change in the stellar SED with metallicity,
through the changing gas-phase abundances, which determine the
temperature and the line emission of the \HII\ regions, through the
opacity of the ISM in the dust, and through the metallicity-dependent
change in grain composition. A full discussion of the effect of
metallicity on the emission line spectra of the \HII\ regions was
given in SED3. In this section we will systematically investigate the
remaining effects.

\begin{figure}
\includegraphics[width=0.9\hsize]{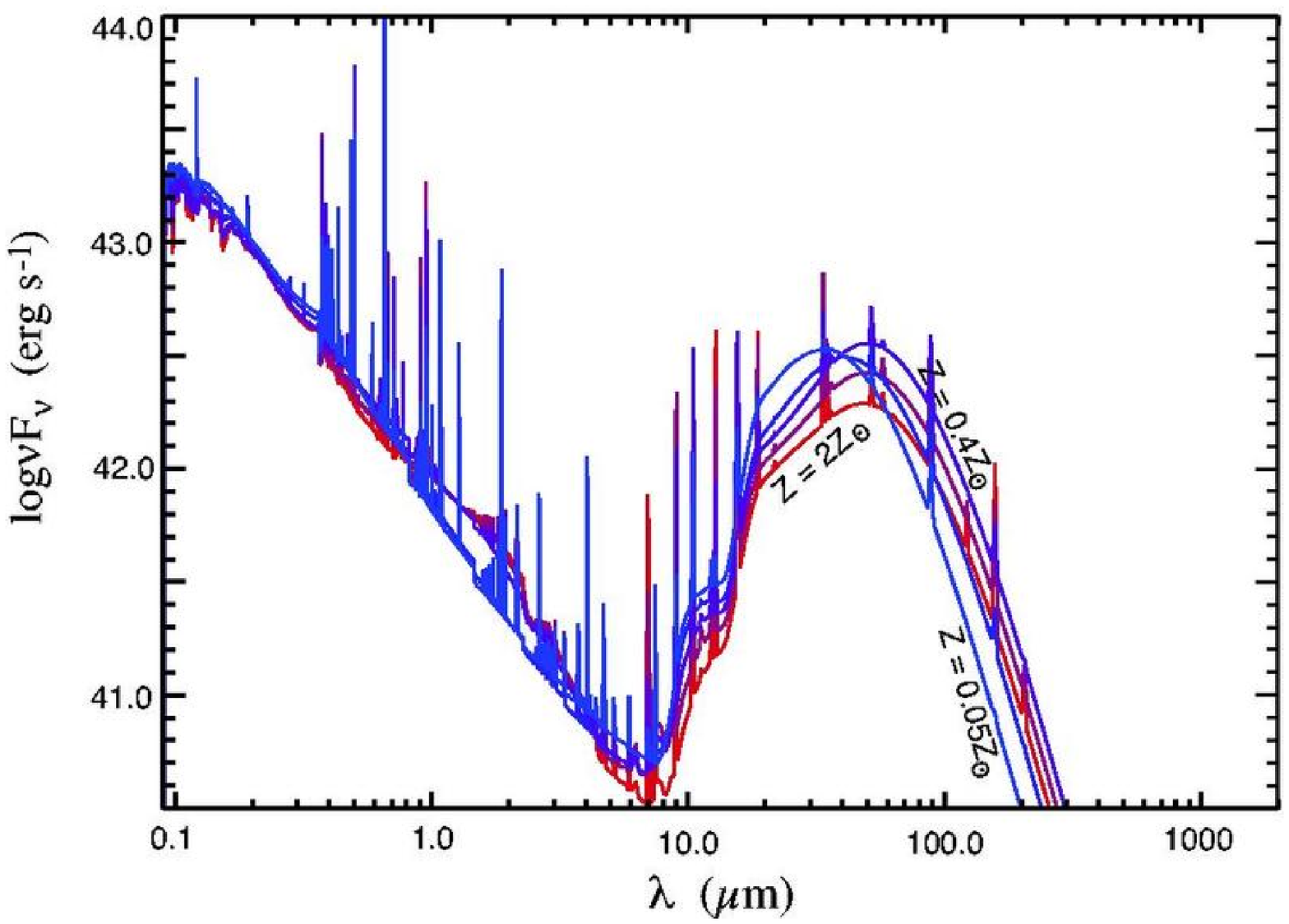}
\includegraphics[width=0.9\hsize]{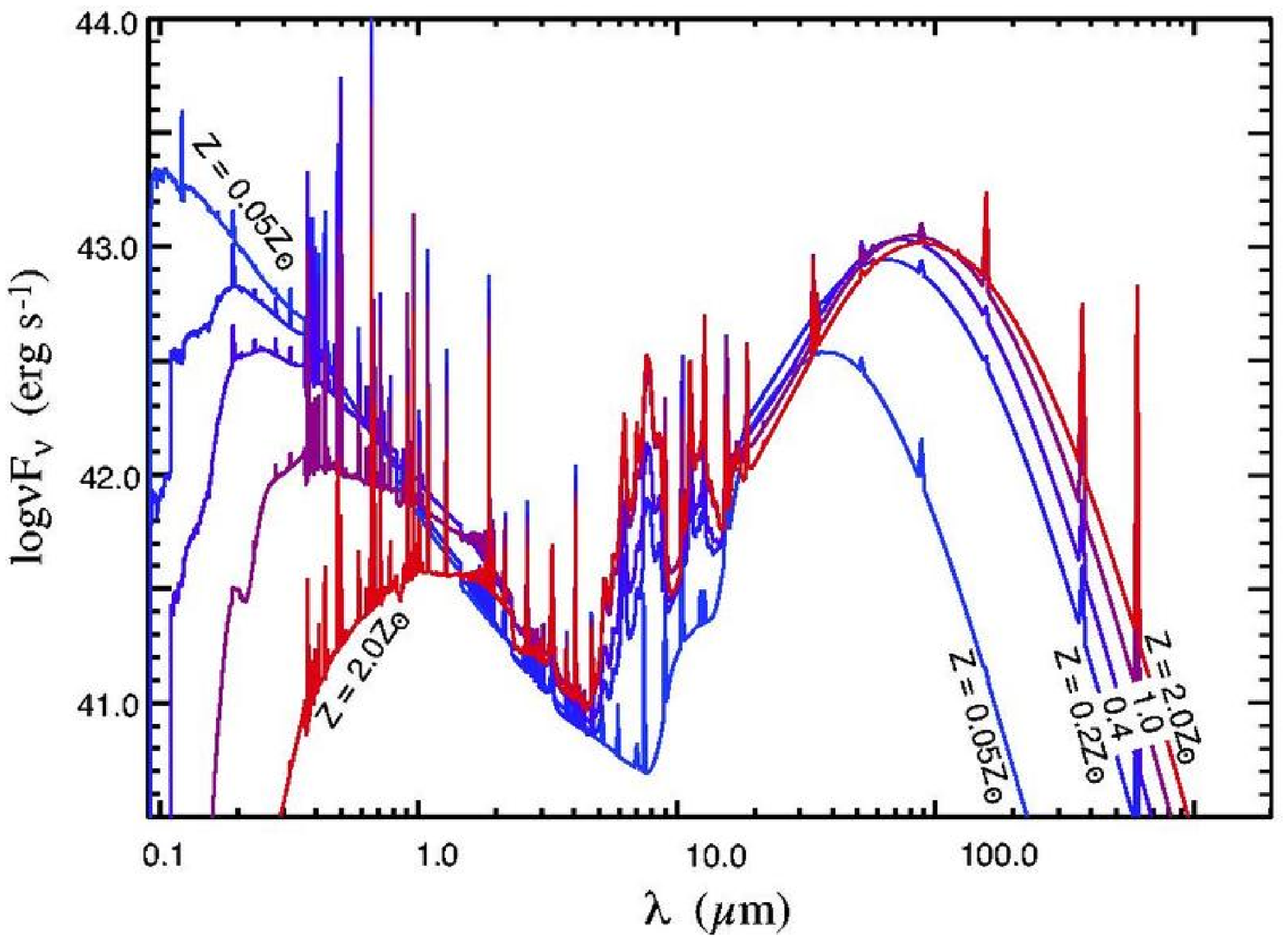}
\caption{The effects of changing metallicity on the \hii\ region
spectra (top) and PDR spectra (bottom) of a log$\Cpar=5$, log$P/k=5$
starburst. }\label{fig:metal}
\end{figure}

In figure \ref{fig:metal} we show our fiducial (\HII\ region \& PDR)
starburst with log $\Cpar=5$ and log$P_{0}/k=5$ computed using the 5
standard Starburst99 metallicities. As the PDR is defined
through a constant column depth of hydrogen, lower metallicity leads
to lower column of dust. This leads to a strong decrease in the
optical-ultraviolet opacity as the metallicity is decreased.  

The metallicity-dependent effects on the \HII\ emission line
spectrum have previously been remarked upon in SED3. 
As the metallicity decreases, the
stellar spectrum becomes harder due to the decreasing opacity of the
stellar atmospheres and winds. In addition, for a given size, a
stellar cluster of lower metallicity has a 
higher ionizing luminosity and lower mechanical luminosity. This leads
to more compact \HII\ regions, and higher ionization parameters in the
surrounding \hii\ region. The fraction of radiation absorbed by the
dust in the \HII\  region depends on the product of metallicity and
ionization parameter \citep{Dopita03}. This product remains
approximately constant with metallicity, ensuring that the flux under
the far-IR bump remains approximately constant for the \HII\ region
SEDs.  

In the PDRs,  the UV to optical SEDs clearly shows the effect of
increasing opacity with metallicity. The far-IR features change in
several ways. First, there is a systematic increase in the relative
strength of the PAH emission features with metallicity. This is a
consequence of the increase in the C/O ratio with metallicity, which
ensures that PAHs account for more absorption at higher metallicity
relative to the silicates, combined with the higher mean dust
temperatures which characterize lower metallicities. Another factor is
the increased strength and hardness of the average radiation field in
the constant column PDR with decreased  
metallicity. As a consequence, the Habing PAH survival criterion 
is only met deeper in the cloud, resulting in overall weaker PAH
emission. 

Such a decrease in the PAH strength with decreasing metallicity has
been observed with both the \emph{Spitzer} and ISO Space Observatories
\citep{Engelbracht05, Rosenberg06, Wu06, Madden06, Jackson07}. However, the
observed depletion of PAHs in the low-metallicity environments may be
even greater than that computed in our models, and both \citet{Wu06}
and \citet{Jackson07} implicate grain destruction processes as acting
more efficiently in the lower metallicity environments. We would hope
that our models, applied to these low-metallicity systems, would be
able to better confirm and quantify this effect. 
 
In the PDRs, the  IR flux can also be seen to increase and become
broader with metallicity up till $Z\sim1.0Z_{\odot}$, after which it
stays approximately constant. This is predominantly due to the
increased dust column. The shift of the IR peak to longer wavelengths
is also partly due to the increased dust column but is also due to the
increasing mechanical luminosity of the starburst with metallicity,
which results in larger  \hii\ regions, with cooler average dust
temperatures.  

\section{Comparison With Observations}\label{sec:Obs}

\subsection{Data Sources}

In order to compare the models with data, we require as close a
homogeneous data-set as available, covering as wide a wavelength range as
possible. For this purpose, we have selected a pair of popular
template starburst galaxies, NGC~6240 and Arp~220 from the 41 ULIRGS
observed with ISO by \citet{Klaas01}. This data set is ideal for our
purpose because their SEDs are well sampled over the full wavelength
regime 1--200\,\mum.   

Flux densities at other wavelengths were collected using the NASA/IPAC
Extragalactic Database (NED) supplemented with a wide selection of
on-line catalogues and papers.  Generally, the UV/optical fluxes are
taken from the third reference catalogue of bright galaxies v3.9
\citep{deVaucouleurs91}. Many optical and near-IR fluxes were taken
from \citet{Spinoglio95}, or from the APM and 2MASS databases. 

The majority of data points in the 1--1300\,\micron\ wavelength range
come from \citet{Klaas97,Klaas01}.  Additional data points were taken
from \citet{Sanders88a}, \citet{Sanders88b}, \citet{Murphy96},
\citet{Rigopoulou96}, \citet{Rigopoulou99}, \citet{SuraceSanders00},
\citet{Lisenfeld00}, \citet{Dunne00,Dunne01}, \citet{Scoville00}, and
\citet{Spoon04} and references therein. 

When available, the UV/optical and near-IR ($JHK$-band) points include
aperture corrections to allow a direct comparison with the larger
aperture mid- and far-IR fluxes (see, for example,
\citet{Spinoglio95}). All UV to near-IR fluxes have been corrected for
Galactic extinction using the E(B$-$V) values based upon IRAS
100\,$\mu$m cirrus emission maps \citep{Schlegel98} and extrapolating
following \citet{Cardelli89}. 

For comparison of the models with a typical \emph{Spitzer} IRS
spectrum of a starburst, we have used the latest (re-calibrated)
version of the spectrum of NGC~7714 from \citet{Brandl06}. 

\subsection{Pan-Spectral Fitting}
Our library of models contains the following elements:
\begin{itemize}
\item{ An ensemble of \HII\ regions surrounding young clusters with
  ages $<10$\ Myr characterized by mean compactness $\cal C$ and
  metallicity $Z$.} 
\item {A set of PDRs surrounding these \HII\ regions with a mean
  geometrical covering factor, $f_{\rm PDR}$.} 
\item{A population of young ($< 1.0$\ Myr) ultra-compact \HII\ regions
  and their PDRs surrounding individual massive stars, characterized
  by a fraction $f_{\rm UCHII}$, where $f_{\rm UCHII}=1.0$ would imply
  that 50\% of massive stars younger than 1.0\ Myr were surrounded by
  ultra-compact \HII\ regions.} 
\item{An older stellar population with ages $10 \leq t \leq
  100$\ Myr. The flux of this component is scaled by a factor $f_{\rm
    old}$, where $f_{\rm old}=1$ would correspond to continuous star
  formation over a total period of 100\ Myr. } 
\item{A foreground turbulent attenuating dust screen, characterized by
  an optical depth in the V- band, $A_{\rm V}$.} 
\item{A re-emission component from the diffuse ISM, characterized by
  the mean Habing field intensity $G_{0}$ and scaled to a percentage
  of the total Bolometric Flux of the Starburst ($\leq 20$\%).} 
\end{itemize}

Ideally, all of these elements should be fitted via a non-linear
least-squares procedure. However, lacking this tool for the time
being, we have elected to make hand-crafted fits to a few template
objects. In addition, we have chosen to make the following simplifying
assumptions: 
\begin{itemize}
\item{We treat the $<10$\ Myr stellar population as an ensemble of
  \HII\ +PDR regions with $f_{\rm PDR}=1.0$. This approximation is
  justified by the observation that the global obscuration of a
  starburst increases with the absolute rate of star formation
  \cite{Buat99,Adelberger00,Dopita02,Vijh03}. This is a natural
  consequence of the  \cite{Kennicutt98} star formation law,
  connecting the surface density of star formation to the surface
  density of gas;  $\Sigma_{\rm SFR} \propto \Sigma_{\rm gas}^n$, with
  $n \sim 1.3-1.6$. Noting that $A_{\rm V} \propto \Sigma_{\rm gas}$,
  it follows that intense starbursts are highly dust-enshrouded.} 
\item{We ignore the contribution of the diffuse dust in the far-IR
  emission. As has already been noted in Section \ref{sec:diffuse},
  the contribution this makes is small, and significant only above 100
  \mum.} 
\end{itemize}

\begin{figure}
\plotone{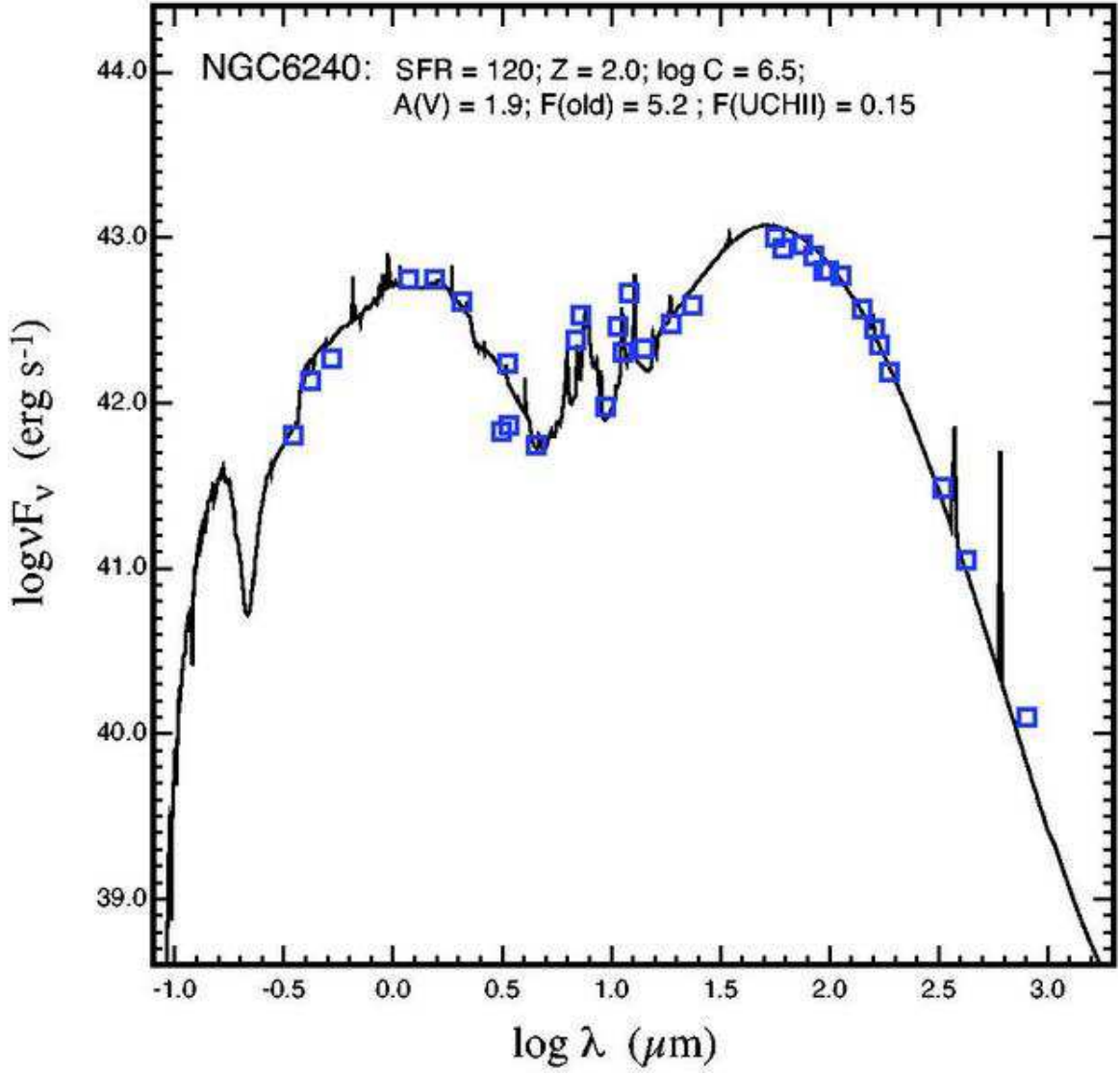}
\caption{The fit to the starburst galaxy NGC~6240. The errors in the
  observed fluxes are similar in magnitude to the height of the blue
  squares. The parameters of the fit are given on the label. Note
  that, although here $A_{\rm V}$ is estimated at 1.9~mag, this
  applies to the older stars. For the younger stars in the \hii\
  regions, $A_{\rm V} \sim 4.4$~mag in this model.}\label{fig:NGC6240} 
\end{figure}

With these assumptions, we show in figure \ref{fig:NGC6240} the fit we
obtain for the galaxy NGC~6240. As can be seen, with five free
parameters we obtain an excellent fit to the observations over nearly
3.5 decades of frequency. In this figure, the data have been scaled to
a star formation rate of 1.0 \Msunpyr\ in order to make this SED
comparable to the others in this paper. The scaling factor required
indicates a total star formation rate of 120 \Msunpyr\ for this
galaxy, a little higher than the value of 102 \Msunpyr\  derived from
the IRAS far-IR luminosity by \citet{Dopita02}. However, the
H$\alpha$ luminosity for this galaxy (also measured by by
\citet{Dopita02}), which comes from the extended gas indicates a star
formation of 14.6 \Msunpyr\ , provided that it is dominated by star
formation, and not by an obscured active nucleus (see below). Since
the far-IR comes from the obscured starburst, and the visible
H$\alpha$ from the regions which are relatively unobscured, these
figures suggest that the total star formation rate is indeed close to
120 \Msunpyr.  

As can be seen in figure \ref{fig:NGC6240}, with a model with only
five free parameters we have obtained an excellent fit to the
observations over nearly 3.5 decades of frequency. However, this
quality of fit to a model which only includes elements of a starburst
system is at first sight extraordinary. NGC~6240 has long been
implicated with an active nucleus, and shows both a strong non-thermal
radio excess, extended radio emission \citep{Gallimore04}, and X-ray
emission associated with a highly dust-obscured AGN
\citep[eg][]{Ikebe00,Risaliti00, Kewley00}. Recently \citet{Armus06}
have directly detected the active nucleus via the [\ion{Ne}{5}] 14.3
\mum\ line using the IRS on the \emph{Spitzer Space Telescope}. From
this, they estimate that the AGN has a flux of $3-5$\% of the
bolometric luminosity. At such low levels, it would account for the
slight excess in the flux in the vicinity of the $6-14$\mum\ PAH
features seen in the observations when compared to our model, but is
not sufficient to compromise the rest of the fit.  

NGC~6240 has an unusually strong contribution from older stars,
suggesting either that the starburst is rather old in this object, or
that the current starburst is the second episode in this galaxy, or
that there is an important contribution of stars older than $\sim
10^8$yrs. All of these hypotheses are consistent with the known status
of NGC~6240 as a post-merger system. 

\begin{figure*}
\plotone{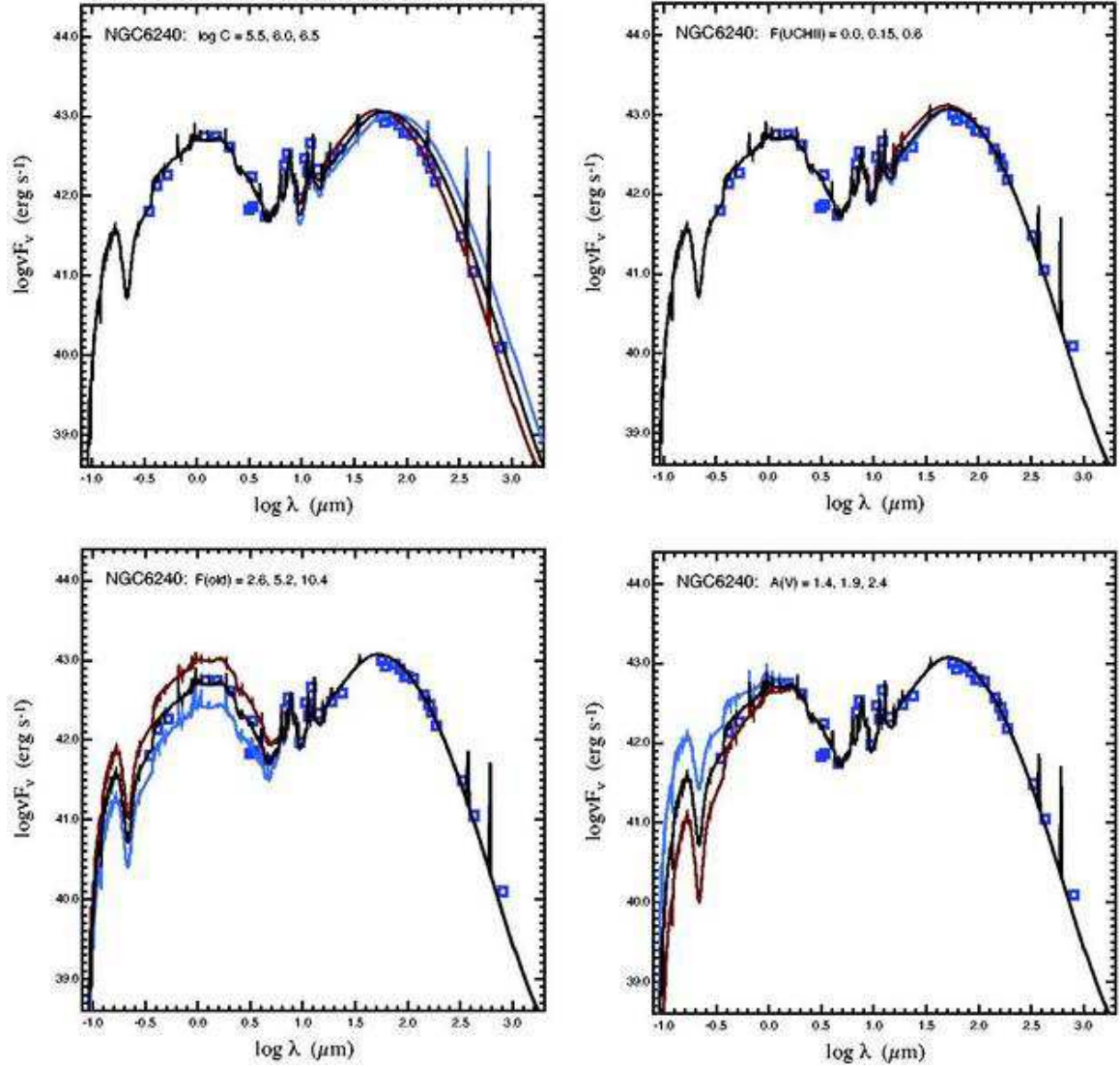}
\caption{The sensitivity of fit to the starburst galaxy NGC~6240 to
  the various parameters. As can be seen, for this galaxy, $\log {\cal
    C}$ is constrained within $\pm0.25$~dex, $A_{\rm V}$ to $\pm
  0.3$~mag,  and $f_{\rm old}$ to within 20\%. The fraction of UCHII
  regions, $f_{\rm UCHII}$ is not very well constrained in this
  galaxy, owing to its high compactness, but would be much better
  constrained in galaxies with $\log {\cal C} < 5$.
}\label{fig:sensfit} 
\end{figure*}

What about the uniqueness of this fit? Fortunately, the different
parameters of the fit act on different parts of the pan-spectral SED,
so it is fairly easy to separate them when we have data covering such
a large range in wavelength. We have performed the test of varying
each of the major parameters, excluding metallicity, systematically around
the best-fit solution, and the result is shown in figure
\ref{fig:sensfit}.  

The effect of varying metallicity is shown in figure \ref{fig:metal},
and has perhaps the most profound effect on the spectrum. To summarize,
the metallicity is most easily determined from the shape and
absorption line intensities of the stellar continuum, and by the
emission line techniques discussed in SED3. It also has an effect on
both the strength of the PAH features and the width of the far-IR
bump. High metallicity gives strong PAH features and wide far-R
bump. In the fitting described in this section we have mostly relied
upon these latter characteristics to estimate the abundance. 

For the remaining parameters, figure  \ref{fig:sensfit} shows that
each affects a different part of the SED. A change in $\log{\cal C}$
only changes the far-IR bump, shifting it in peak wavelength without
appreciable change in the total width. A change in $f_{\rm old}$
simply scales the $0.091 - 5$\mum\ spectrum up and down, leaving the
rest of the SED unchanged. A change in  $A_{\rm V}$ affects the slope
of the visible-UV spectrum. Note that, when $A_{\rm V}$ is higher than
about $5-10$~mag, the 9.7 and 18.0\mum\ silicate absorption features
appear and can be used to constrain the extinction when the the
visible-UV part of the spectrum is too attenuated. The Ultra-Compact
\HII\ region fraction $f_{\rm UCHII}$ mostly affects the $10-30$\mum\
part of the spectrum. In figure \ref{fig:sensfit}, the apparent
sensitivity of the SED to this parameter seems small, but this is
because the inferred  $\log{\cal C}$ for this galaxy is very
high. Thus, the flux of the ordinary \HII\ regions is high in the
$10-100$\mum\ region where the UCHII population is important, and the
contribution of the UCHII regions is veiled. Galaxies with lower
$\log{\cal C}$ show the UCHII contribution much more clearly (see
fig.~\ref{fig:addUCHII}).  

\begin{figure}
\plotone{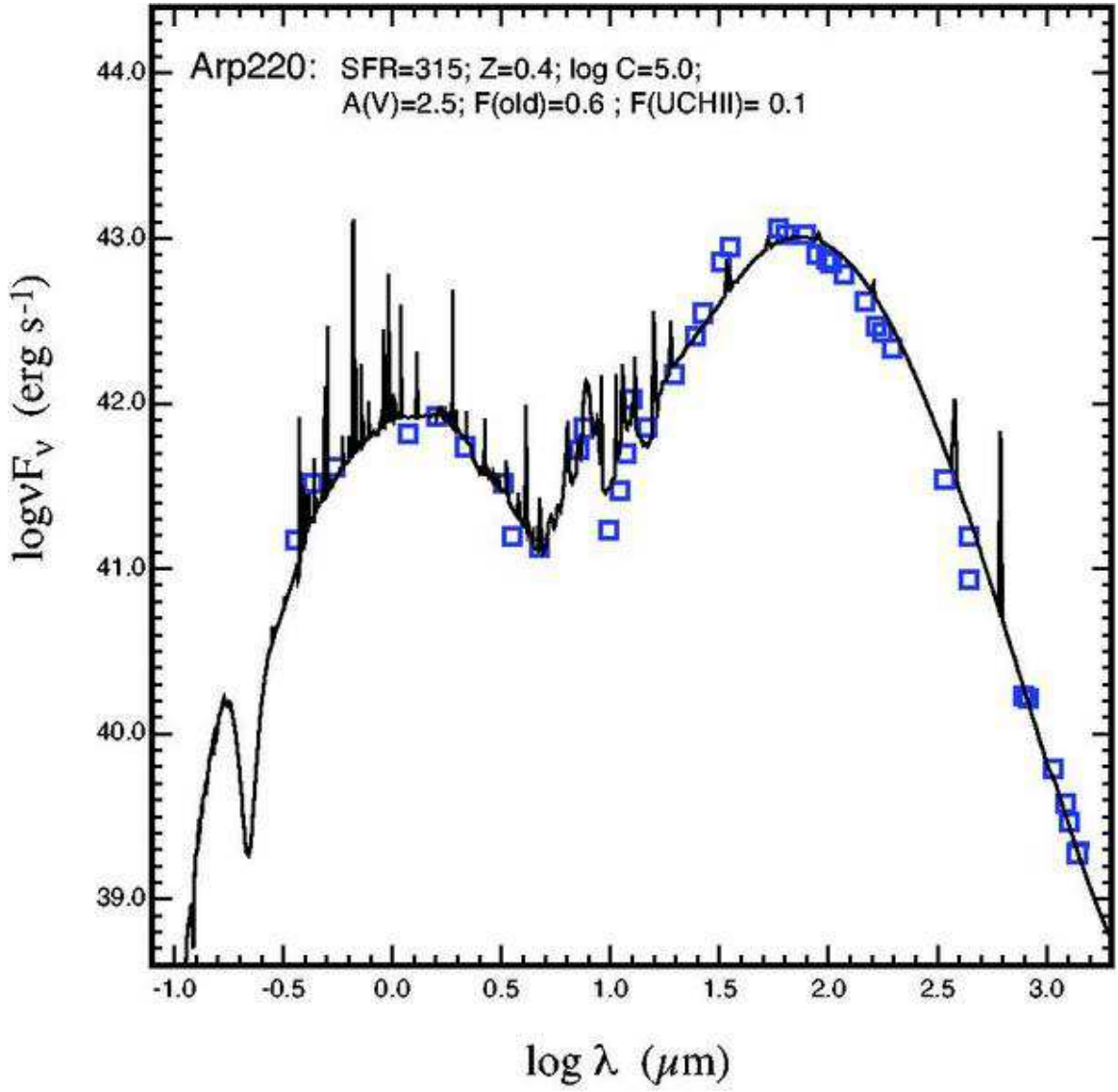}
\caption{The standard fit to the starburst galaxy Arp~220.}\label{fig:Arp220}
\end{figure}

As another example of a famous starburst, we present in
figure \ref{fig:Arp220} our fit to Arp~220, the predominantly
used template of starburst SEDs. This galaxy is characterized by a greater
optical extinction, lower compactness parameter, and a lower
contribution of the old star
population than NGC~6240. In addition, a lower chemical abundance is
indicated by the weaker PAH features, and the narrower, more sharply
peaked far-IR bump. 

Arp~220, as the local ULIRG, has quite often been used as a testbed
for SED modelling. Examples of these can be seen in \citet[][their
figure 9]{Silva98}, \citet[][their figure 8]{Takagi03b} and
\citet[][their figure 5]{Siebenmorgen07} to name a few. While all
these models (including our own) can be seen to fit quite reasonably 
the available
observations of Arp~220, they differ in their model parameters, making
direct comparisons difficult. However the main physical conclusions
drawn from the
models are the same, and comparisons here can give insight into both
the models and Arp~220 itself.

All models suggest a star formation rate (SFR) for Arp~220 
of $\sim$300 \Msunpyr. 
We derive a star formation rate of 315 \Msunpyr, 
which can be compared with the 270 \Msunpyr\  obtained by
\citet{Shioya01}, 260 \Msunpyr\ by \citet{Takagi03b} and 580 \Msunpyr\
by \citet{Silva98}. 

In connection with this is the total luminosity of Arp~220, which is
 $\log L_{*}=12.16 (\lsolar)$ in our models, close to the value of 12.1
of \citet{Takagi03b} and \citet{Siebenmorgen07}, and just below the 12.4
 from \citet{Silva98}. For the total stellar mass of Arp~220
we obtain $\log M_{*}\sim$10.5 (\msolar), higher than the previous SED model
estimates of 10.4  \citep{Silva98} and 10 
\citep{Takagi03b}, but due to the low mass-to-light ratio of our older
stellar component (\S \ref{sec:oldstar}) this value is somewhat
uncertain. All these values indicate that both our model and fit to
Arp~220 are at least consistent with previous models, and suggest the
true values for Arp~220. 

One well known detail of Arp~220 is its very high nuclear extinction; 
$A_{\rm V} \sim 30$ \citep{Shioya01, Spoon04} has been estimated and
even higher estimates from models exist \citep{Siebenmorgen07}. 
This illustrates a
limitation of our fitting procedure. The derived $A_{\rm V}$ is
determined essentially from fitting the attenuation of the older
stellar population, not the nuclear region. When both the 9.7 and
18.0\mum\ silicate absorption features are observed along with optical
or near-IR stellar continuum, we could then modify our fitting
procedure to first fit the $A_{\rm V}$ of a foreground screen for the
\HII\ regions + PDRs implied by the depth of the silicate absorption,
and then apply a second, more optically-thin foreground screen to
match the extinction of the older stars seen in the visible-UV part of
the spectrum.  

\begin{figure}
\plotone{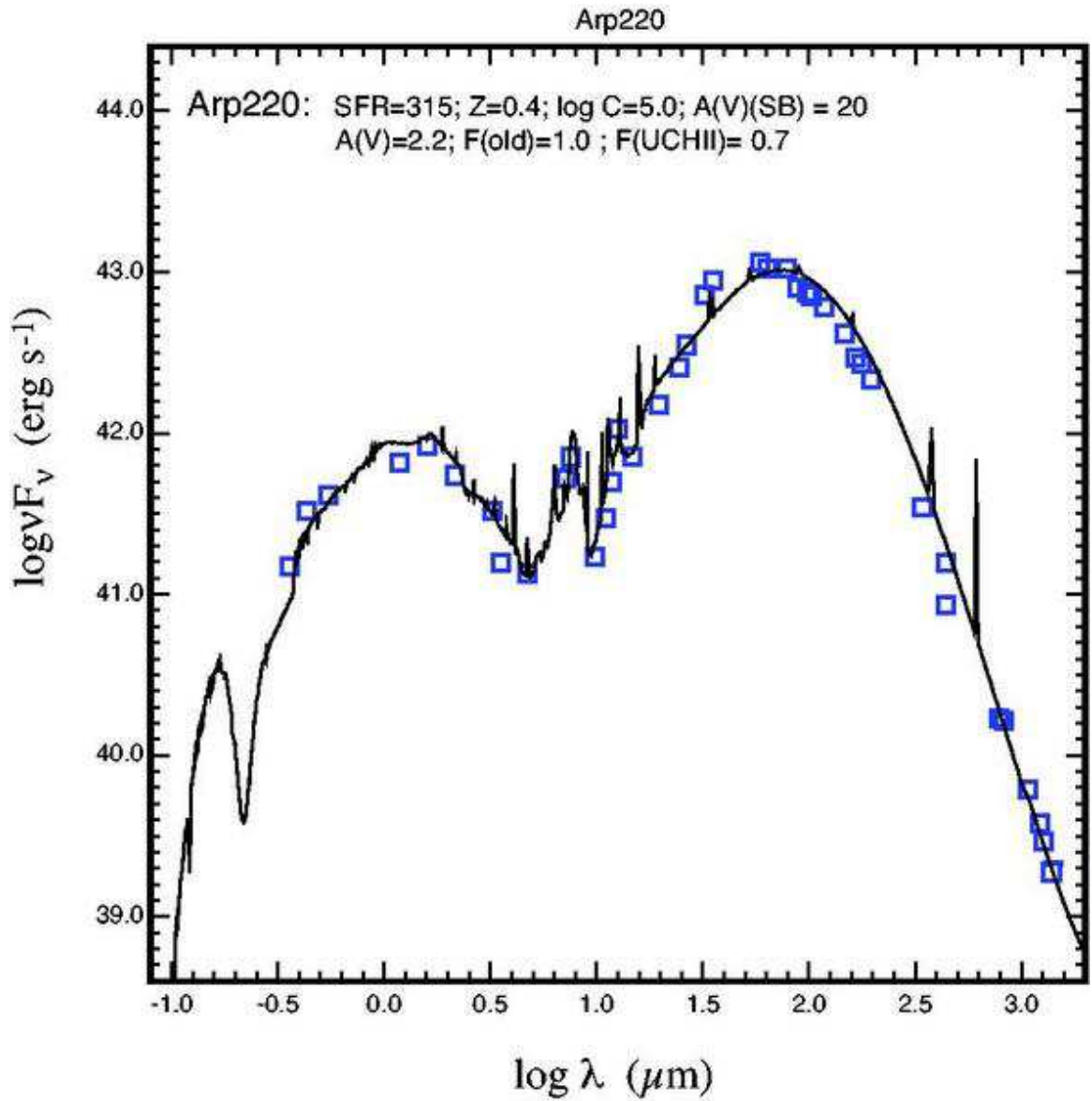}
\caption{The double foreground dust screen fit to the starburst galaxy
  Arp~220. The screen around the starburst nucleus has $A_{\rm V}=
  20$~mag and the diffuse older star component has an attenuation of
  $A_{\rm V}= 2.2$. The fit around the silicate absorption features is
  improved in this more complex model.}\label{fig:Arp220_fit2} 
\end{figure}

In order to test the effect on the fitting parameters we have made
such a model, which we present in figure \ref{fig:Arp220_fit2}. This
model, with $A_{\rm V}= 20$~mag for the starburst component and
$A_{\rm V}= 2.2$ for the older stars certainly fits the region of the
silicate absorption features better, and allows a much larger fraction
of compact \HII\ regions; $f_{\rm UCHII} \sim 0.7$ -  implying that
$\sim 41$\% of all stars younger than 1.0\ Myr are found in compact
\HII\ regions. 

The fact that we derive a lower $A_{\rm V}$ for the starburst than
other authors is not surprising, since the attenuation law that we use
provides greater attenuation in the IR and less in the UV than a
standard extinction law, thanks to the patchy nature of  the
foreground screen. What is surprising is the reduction in the strength
of the emission lines in the visible and UV regions of the
spectrum. This is caused by the much larger nuclear attenuation, and
shows that the equivalent widths of the IR emission lines compared
with the visible or near IR emission lines may be used as a sensitive
diagnostic of nuclear extinction. 

It should be noted that, while Arp~220 is one of the predominantly
used starburst templates, there is increasing evidence that this
object is not a good representative for high-$z$ starforming
galaxies \citep[see eg.][]{Menendez07}. Rather, less extreme objects
such as M82 or NGC 6240 are better local analogs of the high-$z$
actively star-forming objects such as submillimeter galaxies.

\subsection{Fitting Spitzer IRS Spectra}

\begin{figure}
\plotone{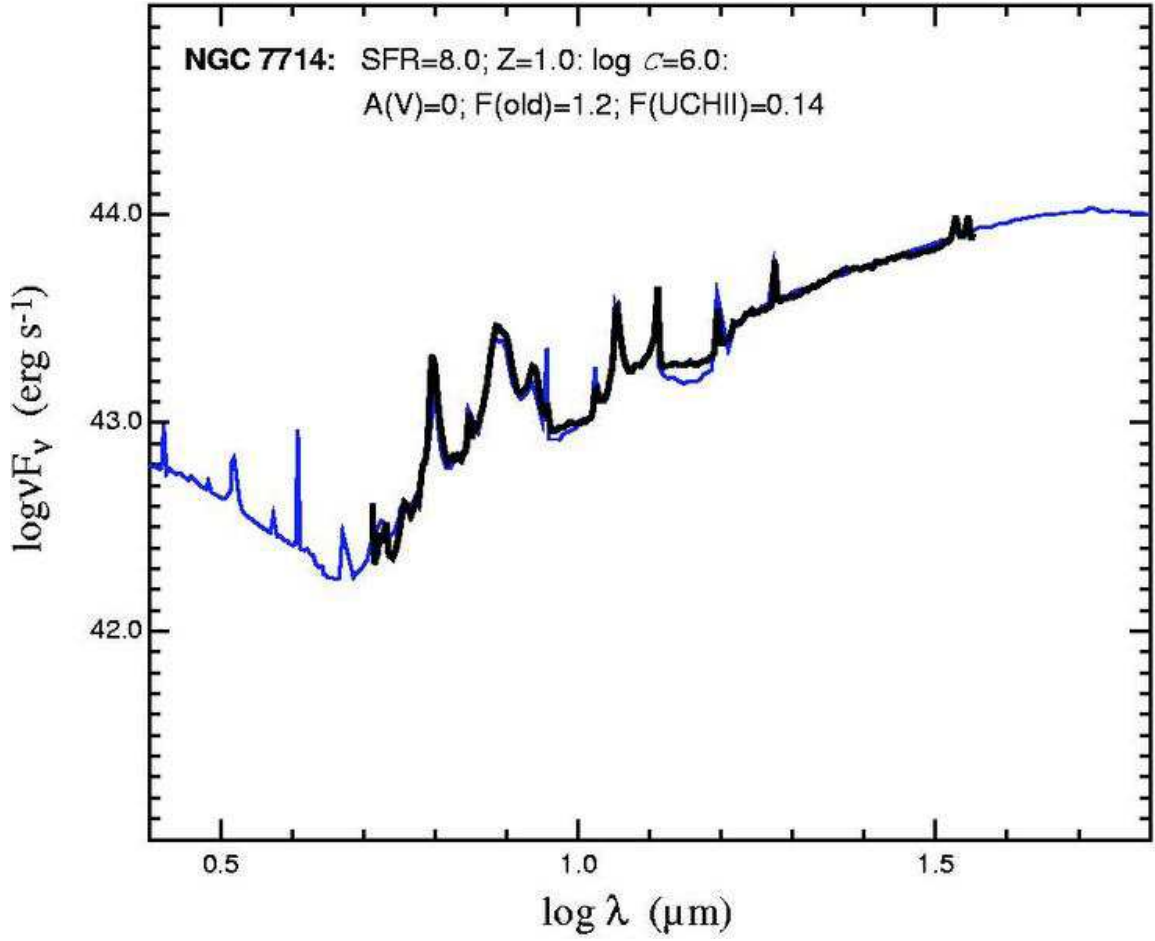}
\caption{The fit to the  \emph{Spitzer Space Observatory} IRS low
  resolution spectra of NGC~7714 from
  \citet{Brandl06}.}\label{fig:NGC7714} 
\end{figure}

As a final example of the fitting process, we compare our fit with the
detailed \emph{Spitzer Space Observatory} IRS low resolution spectra
of NGC~7714 from \citet{Brandl06}. The fit is shown in figure
\ref{fig:NGC7714}. Note that this object has a much lower star
formation rate (8.0 \Msunpyr ) than the previous two examples. The
star formation rate is very well constrained by the normalization
process, and can be determined to an accuracy of better than 5\%,
assuming that the IRS aperture integrates the full extent of the star
forming region.

For this object, the older stellar component is not well-constrained,
as the IRS spectra do not quite extend to short enough wavelengths to
measure it. Also, because the attenuation is not enough to produce
appreciable 9.7\mum\ silicate absorption ($A_{\rm V}< 5$~mag.), we
cannot measure $A_{\rm V}$ from these spectra, so it has been set
equal to zero. Note that the line emission spectrum is quite well
fitted by the model, except for the strength of both the [\ion{S}{4}]
and the [\ion{Ar}{3}] lines in the vicinity of the 9.7\mum\ silicate
absorption band. Again, this may indicate a rather higher obscuration
of the young \HII\ regions than in the model. 

In this spectrum the abundance is fairly well constrained by the
strength of the PAH features and the equivalent width of the emission
lines, while the $20-35$\mum\ slope constrains the values of $\log {\cal
  C}$ and $f_{\rm UCHII}$. The rather steep slope in the continuum
spectrum at about 16\mum\ is a characteristic signature of the
presence of compact \HII\ regions. 

\section{Discussion \& Conclusions}

In this paper we have described an extensive library of pan-spectral
SED models applicable to starburst galaxies, and demonstrated the
promise of these in deriving the physical parameters of
starbursts. These models rely upon a local, rather than a global
solution to the radiative transfer. Such an approach works because of
the fact that, in starburst galaxies, the vast majority of the far-IR
emission arises from absorption of the UV radiation field in a
relatively thin dust layer, the classical photo-dissociation region
(PDR). This region has a typical optical depth corresponding to $A_{\rm V} \sim
3$, and a thickness $\Delta R \sim 300/n_{\rm H}$ pc. In the molecular
regions surrounding normal galactic  \HII\ regions, hydrogen densities
are typically $100 - 1000$ cm$^{-3}$, implying that much of the far-IR
they produce comes from a layer of parsec or sub-parsec
dimensions. In starburst galaxies, interstellar pressures may range up
to a factor of 100 higher than this, producing correspondingly thinner
PDR zones. In addition, the Str{\"o}mgren volume, the volume of the
ionized gas in the \HII\ region surrounding the exciting star or
cluster scales as $n_{\rm H}^{-2}$, making \HII\ regions much more
compact as the pressure in the ISM is increased. 

By simplifying the radiative transfer problem to a local one
connected with individual clusters and their \HII\ regions and PDRs,
we can compute the SED as the sum of a set of effectively independent
components. Our library of models provides the following ingredients
to the pan-spectral SED of starbursts: 
\begin{itemize}
\item{ An ensemble of \HII\ regions surrounding young clusters with
  ages $<10$\ Myr.} 
\item {A set of PDRs surrounding these \HII\ regions.}
\item{A population of young ($< 1.0$\ Myr) ultra-compact \HII\ regions
  and their PDRs surrounding individual massive stars.} 
\item{An older stellar population with ages $10 \leq t \leq 100$\ Myr. }
\item{A foreground turbulent attenuating dust screen. Separate screens
  may be used for the younger $<10$\ Myr population and the older
  stellar population. } 
\item{A re-emission component from the diffuse ISM.} 
\end{itemize}

We have shown that the position of the far-IR dust re-emission peak is
primarily controlled by Compactness Parameter $\cal C$ defined in
section \ref{sec:compact}, although the position and shape of this
feature is also influenced by the mean covering fraction of the PDRs
surrounding the individual \HII\ regions, $f_{\rm PDR}$, investigated
in Section \ref{sec:fpdr}, and by the metallicity discussed in Section
\ref{sec:metal}. In addition we have investigated the effect of the
column density in the PDRs surrounding the \HII\ regions, provided a
global spectrum of an ensemble of compact \HII\ regions derived from
our earlier work, and have investigated the effect of metallicity on
the SED of the older stars with ages $10 \leq t \leq 100$\ Myr. 

Finally, we have provided the attenuation properties of a turbulent
absorbing dusty screen, and have computed simple one-dimensional
models of the thermal emission from the diffuse dust illuminated by
the SED of the older ($>10$\ Myr) population. These models are
characterized by the local diffuse radiation field intensity,
expressed in units of the Habing field intensity, $G_{0}$, where
$G_{0} = 1.0$ corresponds to the intensity of the diffuse radiation
field in the vicinity of the sun. 

We have demonstrated how the far-IR to sub-mm SED is controlled by the
Compactness parameter $\log {\cal C}$, and by the metallicity. This is
of particular application to the high-redshift sub-mm galaxies. As
shown by \cite{Blain04}, a modified Black-Body fit can be made to the
long wavelength side of the far-IR peak in starburst galaxies to
derive a ``dust temperature''. Although the concept of such a dust
temperature is physically meaningless in the light of our models,
which contain a wide distribution in dust temperatures, it is a useful
way to characterize the slope and the position of the sub-mm SED, and
may well be related to the minimum dust temperature in the starburst.  

\cite{Blain04} showed that, in ULIRGs, the ``dust temperature'' derived
in this way is observed to correlate with the absolute luminosity (or
equivalently, to the rate of star formation). In our interpretation we
would conclude that for ULIRGs, the compactness parameter increases
with increasing luminosity. This would be consistent with more
luminous galaxies having greater surface densities of star formation
and greater gas pressures and densities. This in turn is in accord
with the empirical \citet{Kennicutt98} law of star formation;  
$\Sigma_{\rm SFR} \propto \Sigma_{\rm gas}^{1.4\pm0.1}$.

However, the \cite{Blain04} work also showed that the high-redshift
submillimeter selected galaxies (SMGs) provide a similar correlation,
but shifted to higher luminosity. At a given luminosity, the dust
``temperature'' in SMGs is about 20K cooler than in ULIRGs in the local
universe, and at a given dust temperature, the SMGs are typically 30
times as luminous as their ULIRG counterparts.  Given that we have no
reason to suspect lower dust temperatures in sub-mm galaxies, we must
conclude that the starbursts in these galaxies have similar
compactness to local starbursts, but are typically 30 times more
luminous and spatially-extended than local ULIRGs.  

\citet{Takagi03a,Takagi03b} had previously found that most ULIRGS have
a constant surface brightness of order $10^{12}\lsolar$kpc$^{-2}$.
These parameters probably characterize ``maximal'' star formation,
above which gas is blown out into the halo of the galaxy and star
formation quenched. In order to scale the star formation up to the
rates inferred for SMGs ($ \sim 1000-5000 \msolar $yr$^{-1}$), we must
involve a greater area of the galaxy in star formation, rather than
trying to cram more star formation into the same volume. For a typical
value of $10^{13}\lsolar$kpc$^{-2}$, we require ``maximal'' star
formation over an area of $\sim 10$kpc$^{2}$, and the most luminous
SMGs require star formation to be extended over an area of at  order
$\sim 100$kpc$^{2}$.  

From our library of models we have constructed synthetic pan-spectral
SEDs composed of only the following components: 
\begin{itemize}
\item{ The young \HII\ regions with ages $<10$\ Myr characterized by a
  mean compactness, $\cal C$ and metallicity.} 
\item {The PDRs surrounding these \HII\ regions with a mean
  geometrical covering factor, $f_{\rm PDR}$.} 
\item{A population of young ($< 1.0$\ Myr) ultra-compact \HII\ regions
  and their PDRs surrounding individual massive stars, characterized
  by a fraction $f_{\rm UCHII}$, where $f_{\rm UCHII}=1.0$ would imply
  that 50\% of massive stars younger than 1.0\ Myr were surrounded by
  ultra-compact \HII\ regions.} 
\item{The older stellar population with ages $10 \leq t \leq 100$\ Myr,
  scaled by a factor $f_{\rm old}$, where $f_{\rm old}=1$ corresponds
  to continuous star formation over the period of $10 \leq t \leq
  100$\ Myr. } 
\item{A one- or two- component foreground turbulent attenuating dust screen.}
\end{itemize}

Together, these components allow us to construct theoretical
pan-spectral SEDs which encompass the full richness and variety
observed in the SEDs of real starburst galaxies. We note that
different parts of the SED are sensitive in different ways to these
various theoretical components. Therefore, given a sufficient spectral
range in the observations, we can create fits to observed SEDs which
are unique, and which allow us to extract each of the various physical
parameters listed above for the individual starburst
galaxies. Finally, the scaling of the bolometric flux of the
theoretical SED to the observed bolometric flux of the starburst
provides us with an accurate estimate of the total star formation rate
$\dot{M}_*$, measured in \Msunpyr .  

We have demonstrated the utility of this method by fitting theoretical
SEDs to a few famous template starbursts such as Arp~220 and
NGC~6240. These fits have been ``hand-crafted'', and it remains to
automate the process to a multi-variate least-squares fitting
procedure, which we hope to present in a future paper. 

Although this paper has dealt exclusively with starbursts, we note
that many Ultra-Luminous Infrared Galaxies (ULIRGs) are known to
contain an active galactic nucleus. To deal with these cases, we need
to add a further five components to the mix: 
\begin{itemize}
\item{The UV-IR continuum emission of the AGN itself.}
\item{Hot dust from the accretion disk around the AGN, emitting mostly
  in the $5-50$\mum\ waveband.} 
\item {The global emission from the extended Narrow Line Region (NLR)
  surrounding the AGN.} 
\item{A foreground dust screen surrounding the accretion disk, the so
  called ``dusty torus''} 
\item {Non-thermal radio synchrotron emission component or components.}
\end{itemize}

This will be the subject of a future paper in this series, but we note
that many of the required components are already fairly well
understood. The AGN itself is usually approximated by a power-law, or
broken power law, and the NLR component has already been computed by
\citet{Groves06}. The radiative transfer through a dusty torus around
the nucleus is a much more complicated issue that has been dealt with
by several authors. Originally pictured and modelled as a smooth
distribution \citep{Pier92,Granato94,Efstathiou95}, more recent models
assume a more clumpy structure to explain the observed distribution of
the 10\mum\ silicate feature and width of the far-IR peak 
\citep{Nenkova02,Dullemond05,Honig06}. In any case, treatments of the
various components of an AGN SED do exist, and must be considered when
fitting the SEDs of ULIRGs and high-redshift luminous objects. 

The replacement the of semi-empirical approach by the new quantitative
approach to SED fitting of starbursts presented here should greatly
enhance the utility of existing and future surveys, by providing
detailed estimates of the physical parameters of the starbursts. In
this way, it should assist in the derivation of statistical
parameters, demography and cosmic evolution of this class of object,
and should cast light on the nature of starbursts in the high redshift
universe, including sub-mm galaxies and the high-redshift radio
galaxies, when today's massive ``red-and-dead'' Elliptical galaxies
were in the process of assembly, and ULIRGs ruled the star formation
rate density of the Universe. 

\begin{acknowledgements}
M. Dopita acknowledges the support of both the Australian National
University and of  the Australian Research Council (ARC) through his
(2002-2006) ARC Australian Federation Fellowship, and also under the
ARC Discovery projects DP0208445, DP0342844 and DP0664434. WvB
acknowledges support for radio and infrared galaxy studies 
with the Spitzer Space Telescope at UC Merced, including the work
reported here, 
via NASA grants SST GO-1264353, GO-1265551, GO-1279182 and GO-1281587.
 This research has made use of the NASA/IPAC
Extragalactic Database (NED) which is operated by  the Jet Propulsion
Laboratory, California Institute of Technology, under contract with
the National   Aeronautics and Space Administration.  
\end{acknowledgements}

\newpage
\begin{deluxetable}{lll} 
\tabletypesize{\footnotesize}
\tablecaption{ The Solar Abundance Set ($Z_{\odot}$) and 
 logarithmic depletion factors log(D) 
 adopted for each element.\label{tab:abund}}
\tablehead{
\colhead{Element} 
& \colhead{$\log({\rm Z_{\odot}})$}
& \colhead{$\log({\rm D})$}\\
}
\startdata
H & ~0.00 & ~0.00 \\
He & -1.01 & ~0.00 \\ 
C & -3.59 & -0.52 \\
N & -4.20 & -0.22 \\
O & -3.34 & -0.22 \\
Ne & -3.91 & ~0.00 \\
Na & -5.75 & -0.60 \\
Mg & -4.42 & -0.70 \\
Al & -5.61 & -1.70 \\
Si & -4.49 & -1.00 \\
S & -4.79 & -0.22 \\
Cl & -6.40 & -0.30 \\
Ar & -5.44 & ~0.00 \\
Ca & -5.64 & -2.52 \\
Fe & -4.55 & -2.00 \\
Ni & -5.68 & -1.40 \\
\enddata
\end{deluxetable}

\newpage
\begin{deluxetable}{lll} 
\tabletypesize{\footnotesize}
\tablecaption{Normalized Parameters of the Lorentzian Components of the PAH Emission Band
\label{tab:PAHparam}} 
\tablehead{\colhead{$x_0$\tablenotemark{1}} & \colhead{$f_{0}$\tablenotemark{2}} 
& \colhead{$\sigma$\tablenotemark{1}}}
\startdata
3040.3 & 1.006E-04 &  22.4\\
1897.0 & 1.000E-04 &  40.0\\
1754.0 & 1.000E-04 &  40.0\\
1608.5 & 3.420E-04 &  37.8\\
1608.5 & 4.600E-04 &  14.4\\
1593.9 & 2.140E-04 &  34.9\\
1490.0 & 5.000E-05 &  30.0\\
1400.0 & 3.420E-04 & 100.0\\
1313.0 & 1.200E-03 &  28.0\\
1270.0 & 1.280E-03 &  35.0\\
1200.0 & 3.200E-04 &  30.0\\
1163.1 & 8.900E-04 &  27.0\\
 998.0 & 1.630E-04 & 129.1\\
 940.0 & 1.600E-04 &  13.0\\
 890.0 & 1.700E-03 &   4.0\\
 883.0 & 1.800E-03 &  14.1\\
 836.0 & 4.280E-04 &  14.1\\
 813.0 & 2.000E-04 &  15.0\\
 800.0 & 5.300E-04 &  70.7\\
 788.0 & 1.200E-03 &  13.0\\
 737.0 & 3.600E-04 &  18.2\\
 702.0 & 2.570E-04 &  12.9\\
 670.0 & 8.550E-05 &  18.3\\
 635.0 & 1.710E-04 &  18.3\\
 607.0 & 7.800E-04 &   7.5\\
 588.0 & 8.300E-04 &   5.8\\
 576.0 & 4.280E-04 &   4.0\\
 571.0 & 2.140E-04 &  20.0\\
 562.0 & 8.550E-05 &   3.8\\
 530.0 & 1.450E-04 &   7.0\\
\enddata
\tablenotetext{1}{cm$^{-1}$}
\tablenotetext{2}{in normalized units (\fluxnu)}
\end{deluxetable}

\clearpage

\clearpage


\clearpage

\clearpage

\end{document}